\documentclass[9pt,lineno]{article} 
\usepackage{authblk}

\usepackage{lipsum} 
\usepackage[version=4]{mhchem}
\usepackage{siunitx}
\usepackage{graphicx}
\usepackage{pifont}  % http://ctan.org/pkg/pifont
\usepackage{makecell}
\usepackage{colortbl}
\usepackage{hyperref}
\usepackage{subcaption}
\usepackage{svg}
\usepackage{booktabs}
\usepackage{pdflscape}
\usepackage{longtable}
\usepackage{placeins}

\DeclareSIUnit\Molar{M}

\title{Pennsieve: A Collaborative Platform for Translational Neuroscience and Beyond}

\author[1,2]{Zack Goldblum}
\author[1,2]{Zhongchuan Xu}
\author[1,2]{Haoer Shi}
\author[3,4]{Patryk Orzechowski}
\author[1,5]{Jamaal Spence}
\author[1,6]{Kathryn A Davis}
\author[1,2,6]{Brian Litt}
\author[1,2,6]{Nishant Sinha}
\author[3*]{Joost Wagenaar}

\affil[1]{Center for Neuroengineering and Therapeutics, University of Pennsylvania}
\affil[2]{Department of Bioengineering, School of Engineering and Applied Sciences, University of Pennsylvania}
\affil[3]{Department of Biostatistics, Epidemiology and Informatics, University of Pennsylvania}
\affil[4]{Department of Automatics and Robotics, AGH University of Krakow}
\affil[5]{Department of Neuroscience, Perelman School of Medicine, University of Pennsylvania}
\affil[6]{Department of Neurology, Perelman School of Medicine, University of Pennsylvania}
\affil[*]{To whom correspondence should be addressed: joostw@seas.upenn.edu}

\begin{document}

\maketitle
\begin{abstract}
The exponential growth of neuroscientific data necessitates platforms that facilitate data management and multidisciplinary collaboration. In this paper, we introduce Pennsieve — an open-source, cloud-based scientific data management platform built to meet these needs. Pennsieve supports complex multimodal datasets and provides tools for data visualization and analyses. It takes a comprehensive approach to data integration, enabling researchers to define custom metadata schemas and utilize advanced tools to filter and query their data. Pennsieve's modular architecture allows external applications to extend its capabilities, and collaborative workspaces with peer-reviewed data publishing mechanisms promote high-quality datasets optimized for downstream analysis, both in the cloud and on-premises

Pennsieve forms the core for major neuroscience research programs including NIH SPARC Initiative, NIH HEAL Initiative’s PRECISION Human Pain Network, and NIH HEAL RE-JOIN Initiative. It serves more than 80 research groups worldwide, along with several large-scale, inter-institutional projects at clinical sites through the University of Pennsylvania. Underpinning the SPARC.Science, Epilepsy.Science, and Pennsieve Discover portals, Pennsieve stores over 125 TB of scientific data, with 35 TB of data publicly available across more than 350 high-impact datasets. It adheres to the findable, accessible, interoperable, and reusable (FAIR) principles of data sharing and is recognized as one of the NIH-approved Data Repositories. By facilitating scientific data management, discovery, and analysis, Pennsieve fosters a robust and collaborative research ecosystem for neuroscience and beyond.
\end{abstract}

\section{Introduction}
Scientific data management systems are tasked with accommodating the evolving data integration needs of researchers, where detailed metadata and annotations play a crucial role in interpreting complex datasets \cite{Corradi2012A, Amorim2017A, McDougal2016Reproducibility}. This is especially challenging in the neurosciences where the majority of datasets are increasingly multimodal — often spanning multiple domains including neuroimaging, electrophysiology, the electronic medical record, as well as immune and genetic information \cite{Mittal2022Data, NAS2020}. The current neuroscience data landscape is highly fragmented and generally organized into propriety formats and modality-specific archives that do not communicate with each other \cite{martone2024past}. While this provides in-depth capabilities for specific domains, it can inadvertently create data silos that hinder large-scale research efforts  \cite{Hendriks2022Survey, Ascoli2017Win–win, Ferguson2014Big}.  As a consequence, these data repositories are underutilized and their potential for scientific discovery is greatly diminished. Improving the research potential of these repositories will require new approaches to data curation that can ingest and link disparate data to facilitate cross-modal analysis and discovery \cite{glaser2016development}. 

As a field, neuroscience is undergoing a significant transition from primarily closed to open science \cite{martone2024past, anderson2021highlights}. This shift, driven in large part by data sharing policies from NIH the and EU \cite{white2022data}, is part of a broader transformation involving advances in research computing infrastructure and neuroscience-specific data standards \cite{van2021bridging, abrams2022standards, sandstrom2022recommendations}. The resulting exponential increase in neuroscience data volume and complexity presented researchers with unprecedented challenges in data management and cross-institutional collaboration\cite{Cope2021Advances, Kini2016Data, Ament2022The}. As datasets expanded in scale from gigabytes to terabytes and even petabytes \cite{jiang2022petabyte}, traditional methods of data storage, sharing, and analysis became inadequate, particularly for collaborative efforts spanning multiple institutions. Such challenges underscored the need for a standardized approach to data sharing, leading to the widespread adoption of the FAIR (Findable, Accessible, Interoperable, and Reusable) principles \cite{Wilkinson2016The}. The FAIR principles have become a major framework for data interoperability and reproducible research — helping address difficulties in collaborative efforts \cite{Kuplicki2021Common}. 

To further improve the research potential of scientific data, it needs to be standardized, curated with detailed metadata and annotations, and made accessible to both researchers and automated methods \cite{Akil2011Challenges, Martone2004e-Neuroscience:}. These needs call for platforms that can accommodate the scale and complexity of neuroscience data while supporting the collaborative research necessary for discovery and innovation in neuroscience \cite{international2023modular, Sejnowski2014Putting, Biessmann2011Analysis}.

In this paper, we introduce Pennsieve, a scalable, cloud-based platform for modern scientific data management. Pennsieve’s design prioritizes data curation, metadata management, collaborative research, and data publication. Adhering to the FAIR principles of data sharing, Pennsieve ensures that datasets are not only well-managed but also broadly accessible for future research in neuroscience. Below, we will explore current opportunities in neuroscience data management, highlight the landscape of data repositories, and demonstrate how the design of Pennsieve enables a collaborative research ecosystem.

\section{Results}
\subsection{Current Opportunities for Neuroscience Data Management}
Properly managing neuroscience data must addresses the challenges presented by large-scale, diverse datasets. Neuroscience research is becoming increasingly data-driven, a trend accelerated by advances in acquisition techniques and computational capabilities. This provides new opportunities to manage these data in ways that catalyze scientific discovery:

\begin{enumerate}

    \item \textbf{Multimodal Data Management}: Multimodal data enables richer analyses of brain function and behavior by allowing researchers to investigate relationships between different data types \cite{Corradi2012A, Mittal2022Data, hebart2023things}. Effective neuroscience data management systems should facilitate integrating, storing, and analyzing diverse data modalities.

    \item \textbf{Comprehensive Metadata Support}: Integrating metadata and annotations with data is important for understanding and reproducing neuroscience research \cite{Neu2012Practical, Zehl2016Handling}. This makes datasets accurately described, easily searchable, and properly contextualized for reuse by other researchers.
    
    \item \textbf{FAIR Data Sharing}: The FAIR principles ensure that data is well-managed and can be easily shared and reused by others \cite{Wilkinson2016The}. Adopting these principles enables researchers to integrate data from various sources and leverage it for greater discovery \cite{Behan2023FAIR}. 
    
    \item \textbf{Optimizing Data Reliability and Utilization}: Ensuring data quality and making it readily accessible to processing and analysis tools supports reproducibility efforts \cite{McDougal2016Reproducibility, Routier2021Clinica:}. Datasets should be structured and easy to download and use, both through the web and programmatically.  

    \item \textbf{Facilitating Data Integration and Standardization}: The generation of disparate neuroscience datasets complicates data integration and standardization \cite{Sejnowski2014Putting, Neu2012Practical}. When data repositories adopt standardization practices, it improves data interoperability and reuse \cite{Kuplicki2021Common, Bandrowski2012A, Reer2023FAIR}.
    
    \item \textbf{Fostering Collaborative Science}: Isolated data management systems limit collaborative research. Integrated platforms that facilitate data sharing and cross-disciplinary efforts help build a research ecosystem \cite{Cheng2015Going}.

    \item \textbf{Enabling Scalable Analysis}: With scale, data analysis solutions need to become more integrated with data management platforms \cite{sudlow2015uk, hayashi2024brainlife}. No longer is it sufficient to just support data sharing with the intent to enable downloading files.

     \item \textbf{Ensuring Resource Sustainability}: Building a proof of concept is very different from building a lasting resource that people want to use. Sustainable platforms must handle long-term data storage, scale with increasing user engagement, and implement revenue models that recoup operational costs \cite{Wiener2016Enabling, Bach2012A}.

\end{enumerate}

These opportunities are interrelated and should be considered collectively as ways that platforms can contribute to a more robust and efficient research ecosystem. The implementation of FAIR principles underpins efforts to enhance data usability, reliability, and standardization, which in turn facilitates integrating and reusing multimodal datasets. Concurrently, sustainable data resources that support collaboration and scalable analyses enable large-scale research efforts. 

\subsection{Neuroscience Platforms Overview}

Several platforms have been developed to meet the needs of neuroscience researchers, each offering unique features and capabilities tailored to different aspects of data management and analysis. Common across these platforms is a commitment to the FAIR principles and data standardization, though implemented to varying degrees. Challenges remain in comprehensive metadata curation, seamless data integration across modalities, and collaborative research environments. As the volume and complexity of neuroscience data continue to grow, issues of scalability and usability become increasingly important. 

The following survey of prominent neuroscience data repositories showcases a diverse research ecosystem that reflects the breadth of requirements in neuroscience research. Understanding this landscape contextualizes Pennsieve's unique approach to curating data, managing metadata, and accelerating collaborative research.

\subsubsection{Brain-CODE.}  Brain-CODE \cite{Vaccarino2018Brain-CODE:} is a neuroinformatics platform developed by the Ontario Brain Institute (OBI) to manage, analyze, and share multi-dimensional data across different brain conditions. Its federated approach to integrating data allows it to link with other platforms and national databases such as REDCap. It primarily collects neuroimaging, clinical, and multiomics data with Common Data Elements (CDEs) provided by Brain-CODE standardizing data across research projects. The platform leverages a centralized, high-performance computing environment that offers analytical tools and scalable resources — providing researchers with virtual workspaces for data analysis.

\subsubsection{brainlife.io.} brainlife.io \cite{hayashi2024brainlife} is dedicated to reproducible neuroscientific analysis. It combines high-performance computing and cloud resources to run specialized applications for various neuroimaging processing and analysis tasks. These applications can be combined into workflows and used in collaborative research efforts with shared computational resources. Built-in features support reproducible analysis and a standardized 'Datatype' format allows interoperability between applications. With its public-funding model, brainlife.io can work jointly with other public platforms and support their resources. 

\subsubsection{The Data Archive for the BRAIN Initiative (DABI).} DABI \cite{Duncan2023Data:} serves as a specialized repository for human invasive neurophysiology data derived from NIH Brain Research Through Advancing Innovative Neurotechnologies (BRAIN) Initiative projects \cite{insel2013nih}. DABI allows users to upload different data types, including electrophysiology, imaging, pathology, demographic, behavioral data, and surgical notes. While DABI recommends Brain Imaging Data Structure (BIDS) \cite{gorgolewski2016brain} and Neurodata Without Borders (NWB) \cite{TEETERS2015629} formats, it does not strictly require them. The platform offers both centralized and cloud-based storage options, implements data de-identification processes, and provides integrated analytical tools for intracranial EEG (iEEG) data analysis. 

\subsubsection{Distributed Archives for Neurophysiology Data Integration (DANDI).} DANDI \cite{ghosh2019dandi}, also supported by the BRAIN Initiative, is a cloud-based public archive and collaboration platform designed for cellular neurophysiology data sharing, archival, and analysis. DANDI uses the NWB standard as its core data format for interoperability across different neurophysiology data types. The platform offers a JupyterHub web interface for exploring data, and supports data streaming and compute-near-data functionality to work with large datasets in the archive. This programmatic access to the data and metadata encourages secondary research use.

\subsubsection{EBRAINS.} EBRAINS \cite{Appukuttan2021EBRAINS}, developed through the EU-funded Human Brain Project (HBP) \cite{amunts2019human}, is a research infrastructure designed to integrate diverse neuroscience data, tools, and models into a cohesive platform. It gathers data from different modalities and species, hosting not only common data types like MRI and EEG, but also brain atlases, models, and software. There is a significant emphasis on microscopy scans and atlases as well. EBRAINS is compliant with the FAIR principles and has adopted the open Metadata Initiative for Neuroscience Data Structures (openMINDs) framework for metadata management \cite{openMINDS}. The platform aims to improve collaborative brain research by providing tools and resources, including access to high-performance compute.

\subsubsection{The Image and Data Archive (IDA). } The IDA \cite{Crawford2016The} is a platform designed for exploring, archiving, and sharing neuroscience data. Managed by the Laboratory of Neuro Imaging (LONI) at the University of Southern California, the IDA has a global reach, facilitating large-scale, multi-site collaborations. The IDA supports a range of data types, including neuroimaging, biospecimen, and genetic data. It provides integrated tools for data de-identification and search capabilities to create custom data collections. With ongoing support from NIH, among other organizations, the IDA remains an impactful resource for preserving and sharing research data.

\subsubsection{OpenNeuro.} Finally, OpenNeuro \cite{Markiewicz2021The} is a platform that promotes free and open sharing of neuroscience data. Its minimal restrictions on data access and adherence to the FAIR principles enables data reuse by researchers. Initially focused on fMRI data, it has expanded to include other modalities such as EEG, iEEG, MEG, and PET. OpenNeuro only accepts data in the BIDS format to maximize compatibility and uniformity across its database. While the platform itself focuses on managing and sharing data rather than analysis, it partners with other platforms (including brainlife.io) that provide cloud-based tools for data analysis and visualization.

\subsection{Pennsieve Platform}

\begin{figure}[hbt!]
\centering
\includegraphics[width=0.9\textwidth]{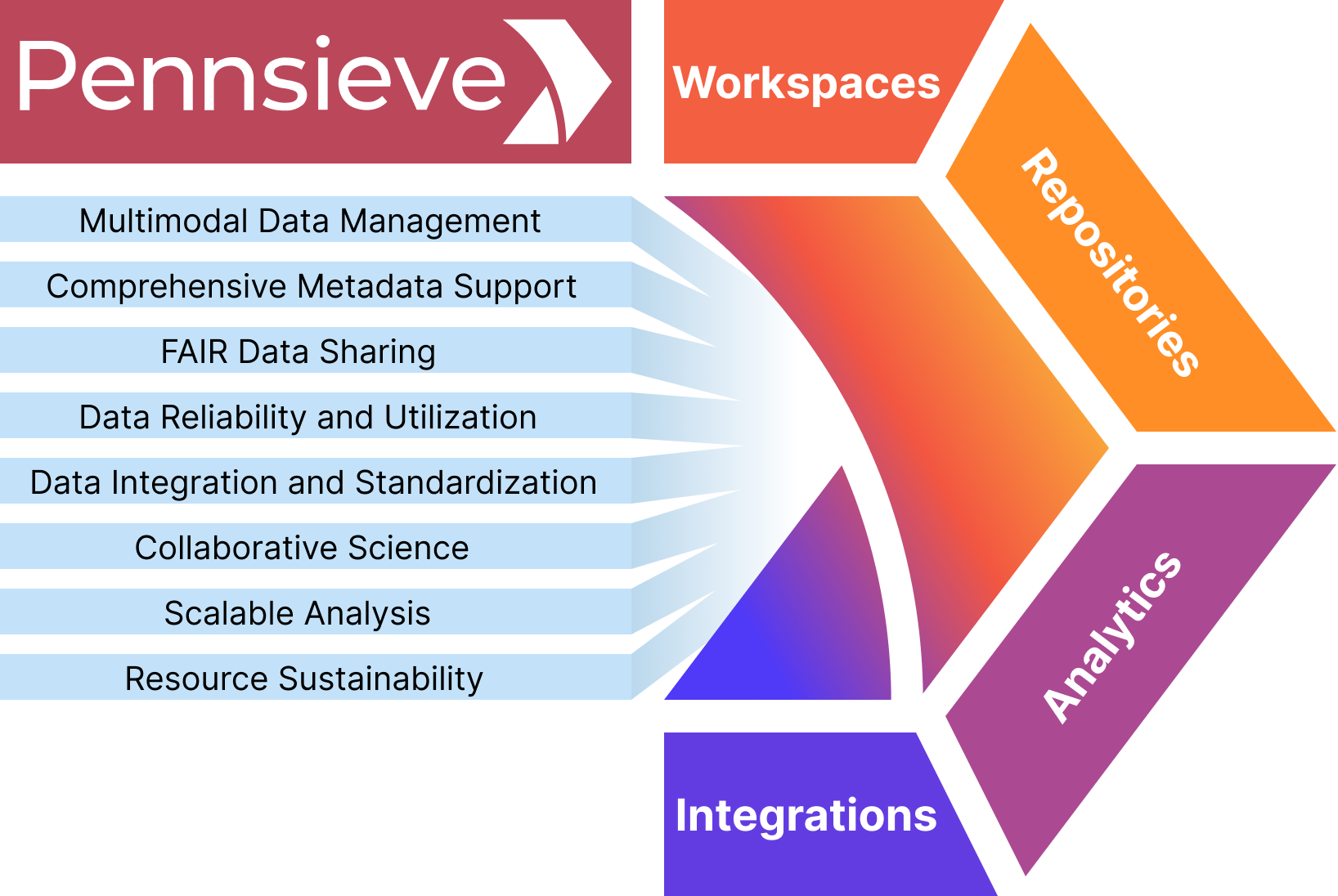}
\caption{The Pennsieve platform serves as a data management infrastructure for the global scientific community. It is built around shared workspaces, data repositories, scalable analytics, and integrations that facilitate collaborative research at scale. \it{FAIR: Findable, Accessible, Interoperable, and Reusable.}}
\label{fig:non-technical_overview}
\end{figure}

Pennsieve is an open-source, cloud-based scientific data management platform (Fig. \ref{fig:non-technical_overview}). It was originally developed for industry under the name Blackfynn Inc. as a cloud-based platform for scientific data management that could support multiple large-scale collaborations. Based on its academic predecessor (IEEG.org) \cite{Kini2016Data}, Blackfynn was funded by DARPA, NIH, and other sources to develop a sustainable, scalable platform for data management and integration for the neurosciences. In 2021, the platform transitioned to an open-source model under the name Pennsieve. It is currently developed and supported as an academic data sharing and integration platform at the University of Pennsylvania.

\begin{figure}[hbt!]
\centering
\includegraphics[width=0.9\textwidth]{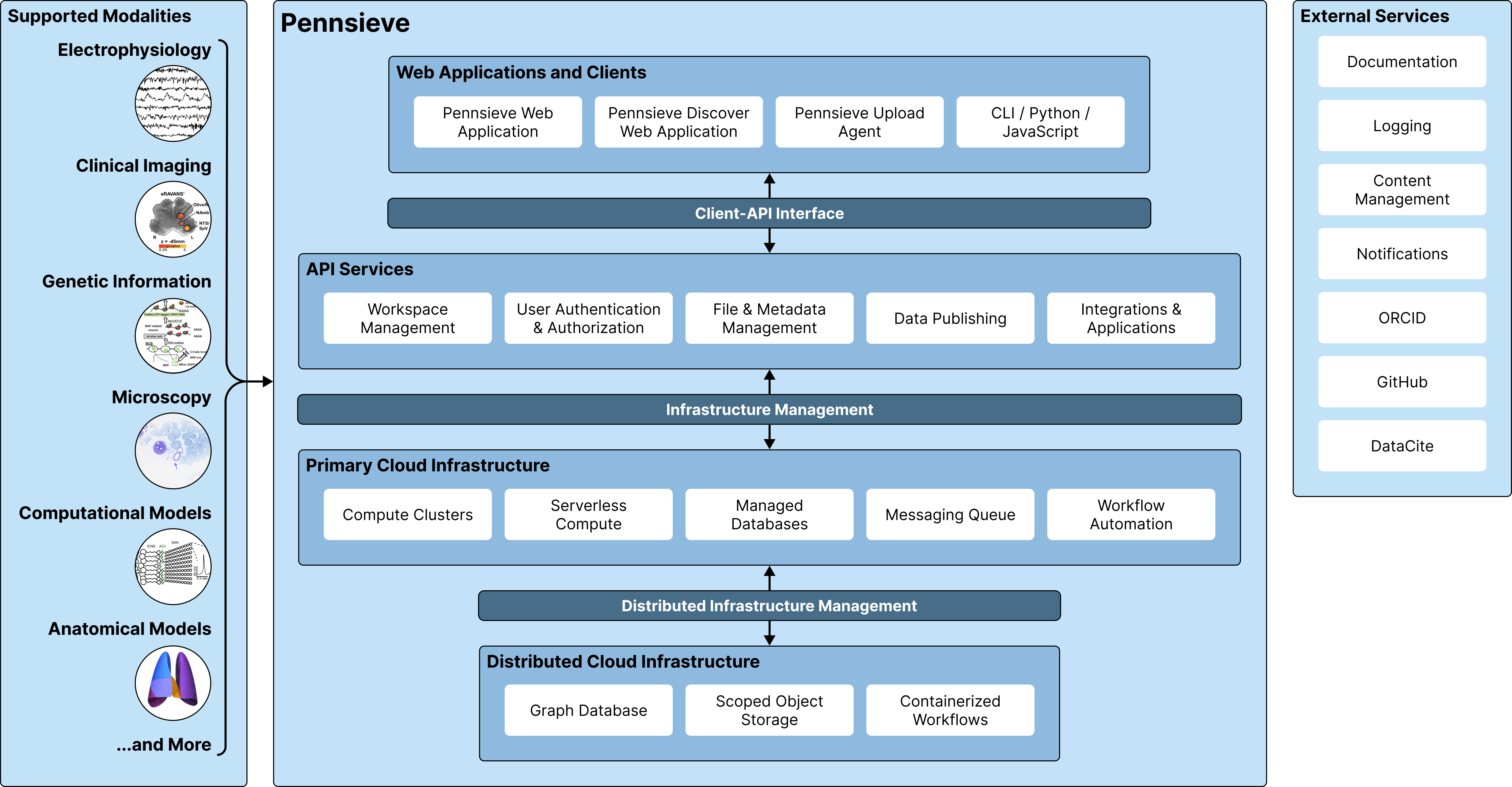}
\caption{A technical overview of the Pennsieve platform. Its multi-tenant architecture supports multiple consortia through dedicated web applications and well-documented APIs.}
\label{fig:overview}
\end{figure}

Pennsieve is a mature platform built on cloud resources that ensure scalability, availability, and security. Fig. \ref{fig:overview} provides an overview of the platform's technical components. The set of features that distinguish Pennsieve are: multimodal data management, flexible metadata schemas, data curation and governance protocols, peer-reviewed data publishing mechanisms, and a scalable, sustainable architecture suited for high-impact scientific research. 

At its core, Pennsieve differentiates between two types of datasets: "private" datasets that are shared within a workspace on the platform, and “published” datasets that are publicly available, versioned, and associated with a Digital Object Identifier (DOI) for citation in publications. Users can belong to multiple workspaces and create datasets that are selectively shared with other workspace members. Several investigator groups leverage this functionality for collaborative efforts while maintaining data privacy. For fully curated datasets ready for publication, Pennsieve has mechanisms that generate versioned snapshots of the dataset that can be made publicly accessible through Pennsieve Discover, the platform's dedicated public data repository. This functionality is used by multiple NIH programs to support FAIR sharing of scientific data. 

In the following sections, we demonstrate how Pennsieve aligns with the previously identified opportunities for neuroscience data management.

\subsubsection{Multimodal Data Management} 
Central to Pennsieve's design is the philosophy that integrating file management with metadata management is crucial for capturing scientific data in its full context. This guiding principle is reflected in the platform's support for diverse data modalities and project structures. These encompass clinical, imaging, timeseries, and molecular data, along with associated metadata. To make full use of this support, users are provided with a suite of features for dataset control, including creation, deletion, file organization, sharing, and permissions management. For programmatic data management, Pennsieve has a rich API that enable operations spanning from data upload and download to complex query and retrieval tasks. Furthermore, datasets can be categorized into collections and assigned status indicators that reflect their progress within publishing pipelines. 

\begin{figure}[hbt!]
\centering
\includegraphics[width=0.9\textwidth]{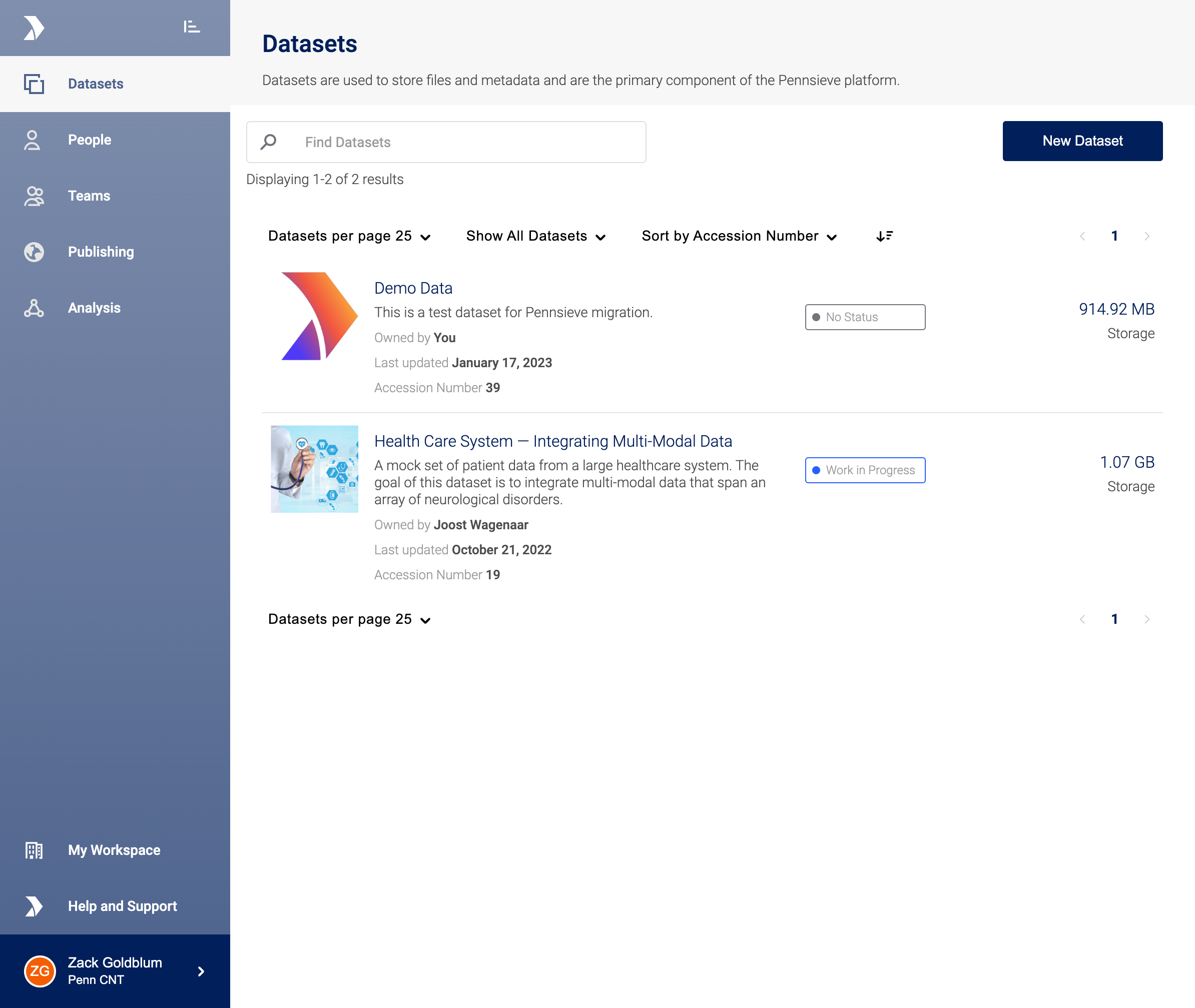}
\caption{The Pennsieve platform's web application interface. This workspace view highlights key features including dataset management, collaborative functionality, publishing workflows, and analysis tools.}
\label{fig:workspace}
\end{figure}

\subsubsection{Comprehensive Metadata Support} 
Pennsieve provides functionality for comprehensive metadata management with an array of tools designed to capture, organize, and visualize them effectively. A Pennsieve dataset consists of both a folder/file structure and a metadata graph which can reference each other. Users can create detailed metadata models with custom schemas to provide context to their data. Using these models, users can establish connections between metadata records and relevant files. To help researchers understand the connections and contextual relationships within their dataset, there are built-in visualizations for metadata graph structures. Metadata is also published alongside files in a dataset, allowing them to be integrated with other systems to facilitate reusing datasets. 

\subsubsection{FAIR Data Sharing}
To promote \textbf{findability}, Pennsieve assigns globally recognized DOIs to each published version of a dataset. These DOIs enable datasets to be cited and indexed through Google Dataset Search and other search engines for discovery by researchers worldwide. They also allow authors of datasets to receive credit when other’s cite their data. Pennsieve Discover is a public repository, closely linked to the platform, that makes published datasets freely \textbf{accessible} to the scientific community. Each dataset published through Pennsieve Discover has a dedicated page that includes details about the dataset and allows users to browse, visualize, and download its files. Datasets are made \textbf{interoperable} by standardizing publishing schemas (the file structure of a dataset) and serializing all metadata into tabulated files. Lastly for \textbf{reusability}, Pennsieve's data publishing mechanisms enforce strict quality standards that ensure datasets are well-documented and ready for reuse. All data, including serialized metadata tables, are exported to cloud object storage where users can fetch and interact with the data without platform restrictions or Pennsieve-specific tools.

Several large-scale, federally-funded programs leverage the Pennsieve platform to support data publication from their investigators. To facilitate high-quality data publications, Pennsieve has implemented several mechanisms to ensure that published datasets meet the requirements for FAIR sharing of data. These include: 1) required data fields such as tags, license, summary, description, and a contributor list, 2) processes for an external data curation team to review datasets prior to publication, and 3) procedures to submit and release new versions of datasets with a complete provenance history.

\subsubsection{Data Reliability and Utilization} 
Data reliability on Pennsieve spans from initial upload to publication and continues through post-publication management. A manifest is created when files are uploaded to a dataset that details the upload status of each file and provides a check for dataset completeness. Once the files are uploaded, Pennsieve maintains a timestamped dataset activity log that attributes any modifications made to the corresponding user. This provides a transparent account of the dataset modification history. Pennsieve's governance mechanisms allow administrators to control who can access, view, edit, or contribute to datasets. For published datasets, post-publication modifications are version controlled, with each dataset version assigned its own DOI. This allows researchers to reference specific versions of data in their work and enables other researchers to access the exact version of the dataset used in a particular study.

Several Pennsieve features help researchers utilize their data. Timeseries and clinical imaging viewers let users visualize and annotate data directly for continuous dataset curation on the platform. Additionally, Pennsieve supports the registration of Compute Nodes that run analytic pipelines. This allows data processing to be performed without needing to export data to external tools. Programmatic access to Pennsieve via an open API enables integrations with other computational workflows and tools. Researchers can use this to automate data processing tasks, utilize Pennsieve data in custom analysis pipelines, and allocate computational resources to handle large-scale datasets. 

\subsubsection{Data Integration and Standardization} 
Pennsieve standardizes data through a combination of protocols and tools that ensure consistency across datasets. The platform supports a range of standardized data formats and conducts validation and quality checks during upload and publication. Datasets are organized in a hierarchical structure that maintains a clear relationship between their different components. Published datasets adhere to a standardized directory structure with pre-defined dataset information fields. Requiring datasets to be consistently formatted and described facilitates discovery and reuse across studies. In addition to these structural protocols, the API can be used to implement standardized data processing and analysis steps. 

\subsubsection{Fostering Collaborative Science} 
The premise of the Pennsieve platform is to foster collaborative science, whether within a single lab, an inter-institutional consortium, or through public access of published datasets. Its multi-tenant architecture allows for independent workspaces that can be used by groups of researchers to organize, curate and privately share datasets (Fig. \ref{fig:workspace}). Datasets are private by default; the owner of the dataset needs to specifically grant other users access to the dataset. Role-based permissions dictate the access and abilities of individual users or teams of users, and datasets can be shared with the entire workspace or only selected teams.

The platform is developed specifically for scientific use-cases and has dedicated functionalities for several data types, such as EEG, clinical imaging, and microscopy imaging. It allows users to view and annotate these files directly in the web application and supports timeseries data streaming. These modality-specific features are key to making the platform more usable for collaborative research without relying on file-sharing and external software packages. 

\subsubsection{Enabling Scalable Analysis}
Pennsieve provides researchers with a suite of tools for analyzing and managing data at scale. Central to this is the Pennsieve agent, a local application that manages large uploads, manifest synchronization, and API interactions. To handle diverse integration needs, Pennsieve-specific libraries have been developed for interacting with this agent and the platform's API via a command line interface (CLI) — available in GO\footnote{https://github.com/Pennsieve/pennsieve-agent}, Python\footnote{https://github.com/Pennsieve/pennsieve-agent-python}, and JavaScript\footnote{https://github.com/Pennsieve/pennsieve-agent-javascript}. Webhooks extend Pennsieve's capabilities by integrating with external applications and automated workflows. 

Building on these established features, Pennsieve is currently advancing its analytics capabilities to enable scientists to execute large-scale analysis workflows directly on the platform. This will allow users to register their own compute resources and mitigate potentially expensive analysis costs by using their existing infrastructure on premises or in the cloud. Additionally, Pennsieve is leveraging its direct GitHub integration to facilitate the creation of analytic workflows from specific repository releases

\subsubsection{Resource Sustainability} 
The Pennsieve platform is developed with sustainability as a core principle. Sustainability requires: 1) the platform to be sufficiently generic to support diverse scientific programs and applications, 2) the implementation of platform functionality with careful consideration of operational costs, 3) the integration of the platform within the larger scientific ecosystem of tools and platforms that research groups use daily, and 4) a balance between novel, cutting-edge functionality and robust, tested technologies expected in a trustworthy platform. 

Beyond platform sustainability, Pennsieve implements both technical and process driven mechanisms for data sustainability. For example, Pennsieve supports distributed data storage on a per-workspace basis. This capability is utilized by individual NIH programs to manage data stores (NIH STRIDES Initiative) separately from the main platform. To reduce resource use, Pennsieve implements staged data life-cycle strategies that minimize storage costs for non-active data

The University of Pennsylvania backs Pennsieve and has committed to store and maintain accessibility of submitted data for a minimum of 10 years. This commitment remains in effect regardless of the active funding status of the Pennsieve project. The long-term goal is to have users pay modest fees for using this resource which are less than they might currently pay for similar, and likely inferior, functionality. Since compute and data download costs are paid by the user, costs for supporting the actual platform are moderate and we are considering subscription or usage models that distribute these costs equitably. 

As an open-source platform, Pennsieve has operational transparency and encourages contributions from developers and researchers. This allows the platform to evolve in response to the community’s needs — informing design decisions that enhance its functionality and appeal. Detailed documentation and guides, a user-friendly interface, extensibility through integrations, and collaborative data management contribute to Pennsieve’s enduring utility to the scientific community.

\subsection{Pennsieve's Impact}
\subsubsection{Building a Collaborative Research Ecosystem}
\begin{figure}[hbt!]
\centering
\includegraphics[width=0.8\textwidth]{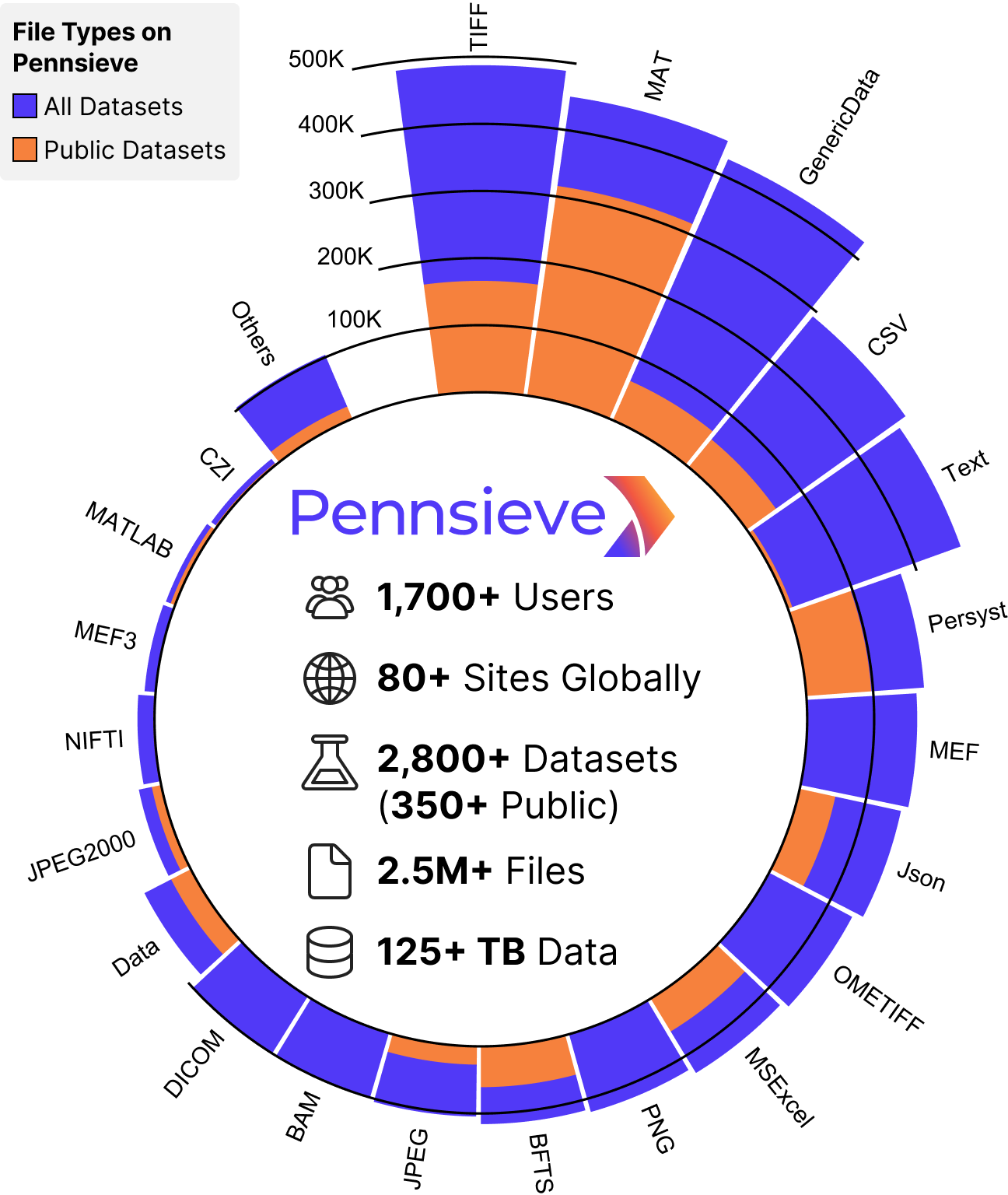}
\caption{Pennsieve platform metrics and file type distribution. The circular chart illustrates the variety of file types supported, with comparative proportions for all datasets and public datasets.}
\label{fig:infographic}
\end{figure}

Pennsieve hosts a vast and diverse collection of scientific datasets amounting to over 2.5 million files and 125 TB of data, which is growing daily (Fig. \ref{fig:infographic}). The repository supports research efforts ranging from individual investigator groups to large-scale, NIH-funded consortia. To date, more than 350 datasets have been made publicly available through Pennsieve Discover. These span from small collections (\textless  10 GB) to large datasets (\textgreater  5 TB) with complex data structures and metadata graphs.

The platform's adoption at leading research institutions exemplifies its impact on collaborative science. The Center for Neuroengineering and Therapeutics (CNT) at the University of Pennsylvania utilizes Pennsieve to lead inter-institutional projects, aggregating data and sharing analyses across clinical sites. It has also enabled the CNT to implement standardized epilepsy coregistration pipelines that significantly reduce overhead for data analysis. Similarly, the Penn Immune Health Project (I3H) leverages Pennsieve to automate and scale their analytic workflows, including the generation of reports for cell classification and immune profiling.

\subsubsection{Supporting Major Research Initiatives}
In support of a collaborative research ecosystem, Pennsieve provides data management infrastructure for broader scientific repositories and projects. The includes several NIH initiatives: the Stimulating Peripheral Activity to Relieve Conditions (SPARC) program \cite{SPARC2024}, the Helping to End Addiction Long-term (HEAL) Initiative's \cite{HEAL2024} Restoring Joint Health and Function to Reduce Pain (RE-JOIN) Consortium, and the Program to Reveal and Evaluate Cells-to-gene Information that Specify Intricacies, Origins, and the Nature of Human Pain (PRECISION Human Pain) network. These initiatives, and more than 80 research groups worldwide, use Pennsieve for data submission, curation, and publication.

Pennsieve underpins several domain-specific portals that promote data accessibility in specialized research areas. The SPARC.Science portal serves as a central hub for datasets from the SPARC program, HEAL RE-JOIN Consortium, and HEAL PRECISION Human Pain network. It contains over 90 active projects, 220 datasets, 38 anatomical models, and 45 computational models focused on peripheral nervous system analysis. As a HEAL-compliant repository, it also accepts datasets from other HEAL initiatives and NIH efforts within the scope of pain and addiction research. 

Pennsieve is also expanding its impact on epilepsy research through the Epilepsy.Science portal. This project brings together Pennsieve’s scalable data management, the Brain Data Science Platform (BDSP)'s extensive data resources and analytic tools \cite{BDSP2024}, and cloud credits from the Amazon Web Services (AWS) Open Data Sponsorship Program to sustain public access. The aim is to develop an extensive catalog of high-quality, multimodal epilepsy datasets from broad contexts for research and clinical translation.

Other specialized efforts are supported by Pennsieve. For example, the Pediatric Quantitative EEG Strategic Taskforce (PedQuEST), centered on advancing brain-focused care for critically ill children using quantitative EEG, and the Childhood Status Epilepticus and Epilepsy Determinants of Outcome (SEED) project, a extensive study on childhood status epilepticus in Nigeria. 

The use of Pennsieve infrastructure to support these efforts demonstrates its utility beyond traditional neuroscience data management. Its capacity to handle large-scale, complex datasets makes it an impactful platform for various NIH programs and other targeted projects.

\section{Discussion}
Pennsieve offers a comprehensive neuroscience data management platform that addresses many of the field's current challenges. Its emphasis on metadata curation, collaboration, data publishing, and extensibility positions it well to support the demands of modern neuroscience research. As the field continues to generate increasingly large and complex datasets, platforms like Pennsieve will play an important role in catalyzing scientific insights in a variety of fields. Though current major efforts are focused on immune health and the brain and nervous system, Pennsieve’s reach is increasing daily, and there is tremendous opportunity to leverage its open-source model in reaching other fields, applications, and disciplines.  

Pennsieve's unique approach focuses not only on making data publicly available but also on \textit{how} datasets are presented and shared. Visually appealing and informative landing pages for public datasets aim to capture researchers' interest and facilitate meaningful connections between data contributors and potential collaborators. Moreover, Pennsieve supports internal data sharing between collaborators within and across institutions, integrating into existing scientific workflows. This lowers the barrier for researchers to make their data publicly available, aligning with open science initiatives and data sharing mandates. 

Pennsieve's transition from industry to academia was pivotal, as it reshaped the platform into a sustainability model that prioritizes open science and the FAIR principles. Consequently, Pennsieve gained credibility and adoption among researchers, with increased engagement from open-source initiatives. This transition from funded development to a sustainable operational model was challenging, and the constant effort to balance the immediate needs of supported projects with long-term sustainability considerations necessitated compromises. For instance, making large-scale data accessible at cost required a reduced focus on supporting analysis directly on the platform. Choices such as this reflect Pennsieve's commitment to creating a sustainable infrastructure that can continue to serve both the neuroscience and broader scientific communities beyond initial funding phases.

\subsection{Future Directions}
While Pennsieve offers comprehensive infrastructure for data management and collaboration, it is still in active development, with a roadmap for new functionality spanning the coming years. We are increasingly focused on supporting sustainable, distributed analytics through the platform, and expect that these capabilities will increase Pennsieve's value to researchers beyond data management and publication. In addition, we are actively pursuing integrations with other platforms - both academic and commercial - to expand Pennsieve's impact within the larger ecosystem of scientific tools and infrastructure available to the neuroscience community. We strongly believe that the next 10 years of neuroscience research will trend towards the integration of analytic tools, repositories, and other efforts that systematically increase our ability to leverage data for new discoveries and the development of cures for patients with disease. 

Our vision for Pennsieve is inextricably tied to the evolution of data platforms and utilities across multiple sectors. As science, business, education, and government grapple with exponentially expanding data, platforms like ours promise to address some of society's most vexing problems. Sustainability, as mentioned earlier, remains a key concern. We firmly believe that our platform and others will remain self-sustaining if they continue to provide real value to their users.

At the same time, it's vital to acknowledge that no single platform will dominate this space. The future lies in linking platforms across groups and fields, and sharing the most valuable tools between them. Several challenges persist in this domain: addressing the "data lifecycle", navigating privacy concerns and data de-identification, balancing data sharing across institutions while maintaining individual control and ownership of resources, and reconciling the revenue focus of for-profit entities with societal needs to leverage data, reduce costs, and improve the human condition. 

These challenges are not unique to data platforms and sharing in the biological sciences; they extend to many activities affecting our daily lives. While the ideas presented here are a start to this process, we anticipate more complex discussions as relevant technologies advance in complexity and societal uptake increases.

\section{Acknowledgments}
Pennsieve, formerly known as Blackfynn, was built over several years, and received funding from multiple sources, including foundations, Michael J. Fox Foundation, Epilepsy Foundation, CURE Epilepsy; institutions, Penn Center for Neuroengineering and Therapeutics, Penn Institute for Immunology \& Immune Health, Penn Institute for Biomedical Informatics, and Children's Hospital of Philadelphia; as well as federal funding from National Institutes of Health, grants no. 1U24NS135547, 1U24NS134536, OT3OD025347, OT3OD025347, UC4DK112217, R44DA044929, K01ES025436, and U24NS063930; and Defense Advanced Research Projects Agency, grants no. DARPA-D2-1801, and DARPA-BAA-15-35. N.S. received funding from the National Institute of Neurological Disorders and Stroke (NINDS) of the National Institutes of Health, awards no. K99NS138680, R01NS116504, and R01NS125137, as well as the Department of Defense, grant no. W81XWH2210593. In particular, Pennsieve would like to acknowledge leadership at NINDS and NIH Common Fund: Walter Korshetz, Vicki Whittemore, Randall Stewart, Andrew Weitz, Felicia Qashu, and others, who seeded the vision for Pennsieve and its predecessor, IEEG.org, through their funding, guidance, and encouragement. 

\bibliography{references}
\bibliographystyle{naturemag} %Nature Communication

\clearpage
\setcounter{page}{1}
\appendix

\section*{Supplementary Material}
\renewcommand\thefigure{\thesection.\arabic{figure}}    
\setcounter{figure}{0}

\section{Pennsieve Platform}
To access the platform, visit \href{https://app.pennsieve.io/}{\textbf{\textit{Pennsieve}}}. For public datasets, explore \href{https://discover.pennsieve.io/}{\textbf{\textit{Pennsieve Discover}}}. If you would like to publish your data on Pennsieve, please reference the \href{https://docs.pennsieve.io/docs/overview#data-publishing-workflow-schematic}{\textbf{publishing documentation}}. Extensive documentation for Pennsieve can be found in the \href{https://docs.pennsieve.io}{\textbf{Documentation Hub}} and the \href{https://docs.pennsieve.io/reference}{\textbf{API reference}} provides code snippets and tutorials for interacting with the Pennsieve API in more than 20 programming languages. The fully open-source code for the platform can be found on \href{https://github.com/Pennsieve}{\textbf{\textit{GitHub}}}.

\begin{figure}[hbt!]
\centering
\includegraphics[width=0.9\textwidth]{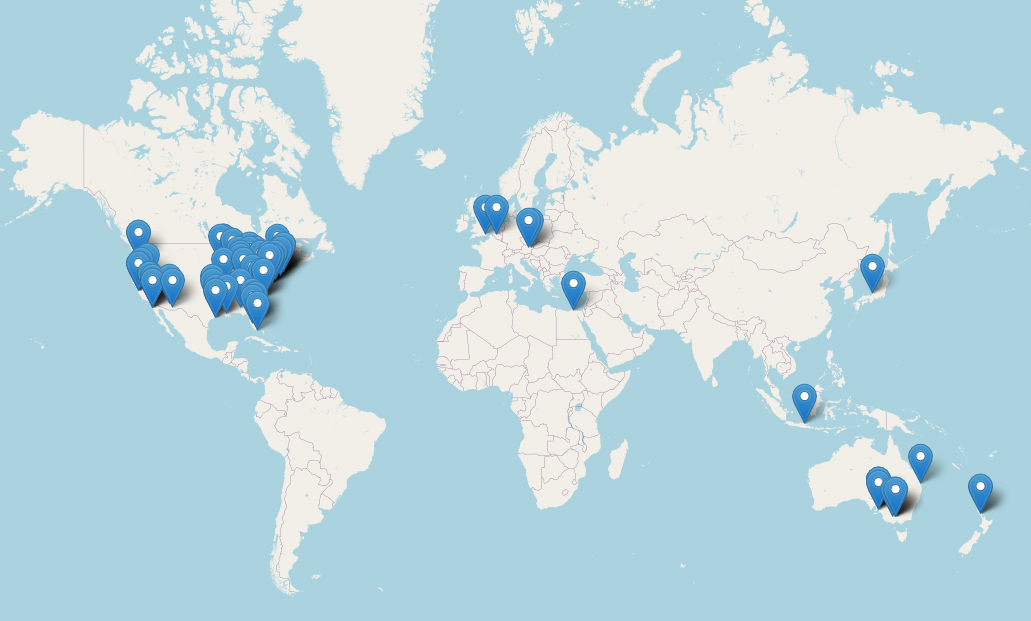}
\caption{Pennsieve is utilized by more than 1,700 users across 80+ research sites worldwide.}
\label{fig:map}
\end{figure}

At the time of writing, Pennsieve has over 1,700 registered users from more than 80 different research sites (Fig. \ref{fig:map}). With more than 350 public datasets (2,800+ public and private combined), it stores over 2.5 million files comprising 125 TB of data. This makes Pennsieve one of the largest, fully-maintained data resources for the neuroscience community.

\begin{figure}[hbt!]
\centering
\includegraphics[width=0.9\textwidth]{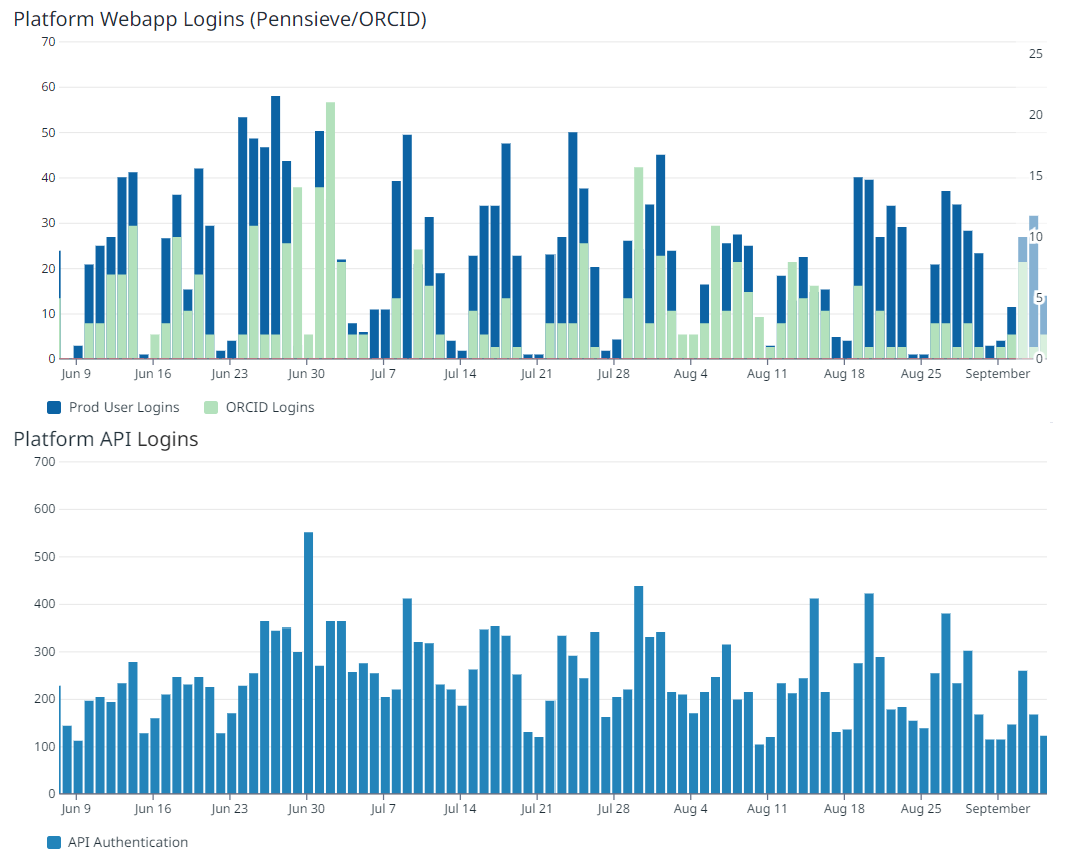}
\caption{Daily user engagement with Pennsieve. \textit{Top:} Web application logins, distinguishing between standard user logins and ORCID authentications. \textit{Bottom:} API authentication requests, indicating programmatic platform usage.}
\label{fig:logins}
\end{figure}

\begin{figure}[hbt!]
\centering
\includegraphics[width=0.9\textwidth]{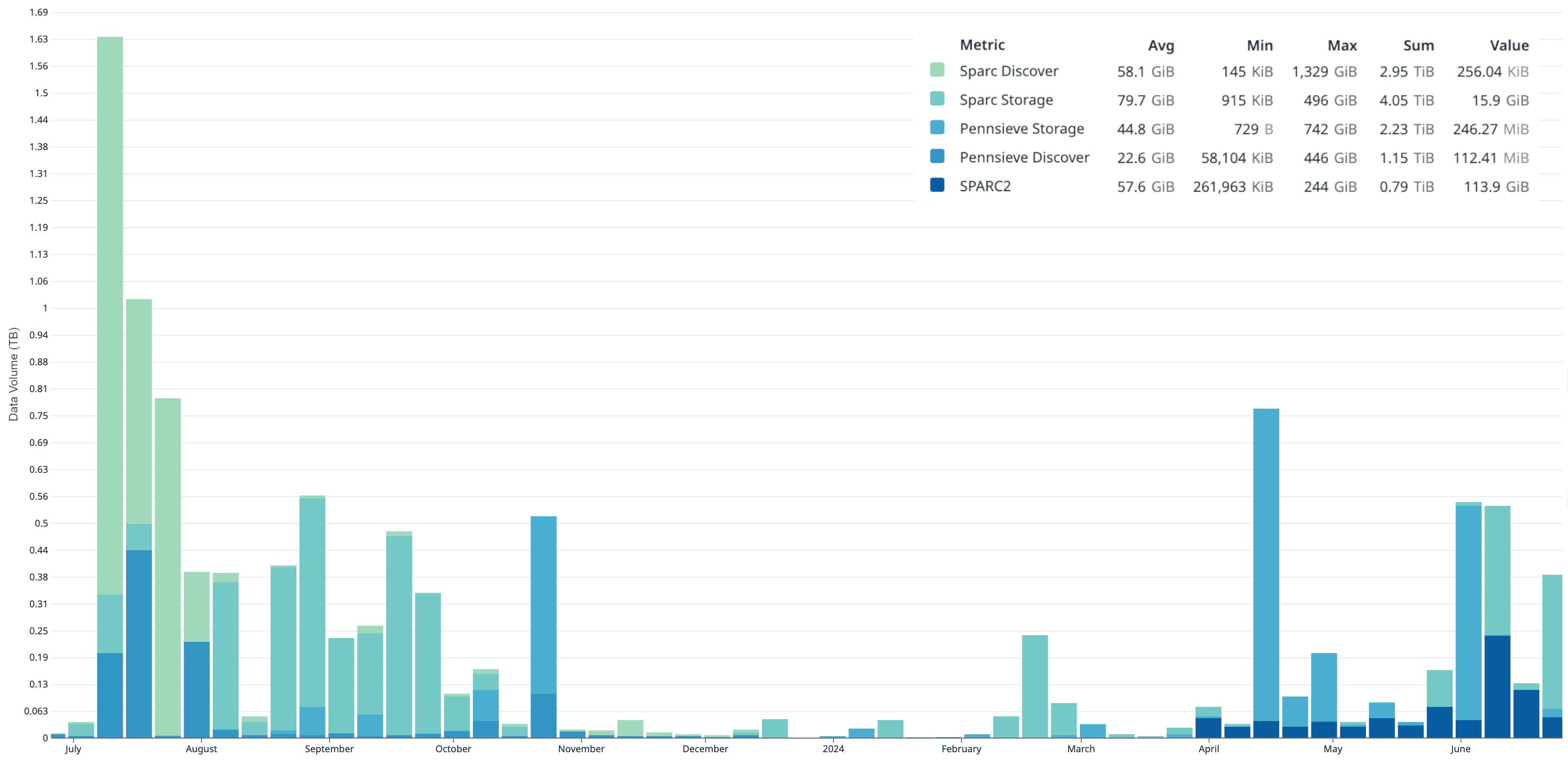}
\caption{Monthly data volume (in TB) downloaded from Pennsieve and related services. The stacked bar chart shows the distribution across different services. Public services (Pennsieve Discover and SPARC Discover) account for a significant portion of the transfers.}
\label{fig:downloads}
\end{figure}

Over 100 users use Pennsieve daily as their repository of choice for storing and managing scientific data (Fig. \ref{fig:logins}). On average, 80+ GB of data is downloaded weekly from Pennsieve's public services (Fig. \ref{fig:downloads}).

\subsection{Data Management Workflow}
\begin{figure}[!htbp]
\centering
\includegraphics[width=0.9\textwidth]{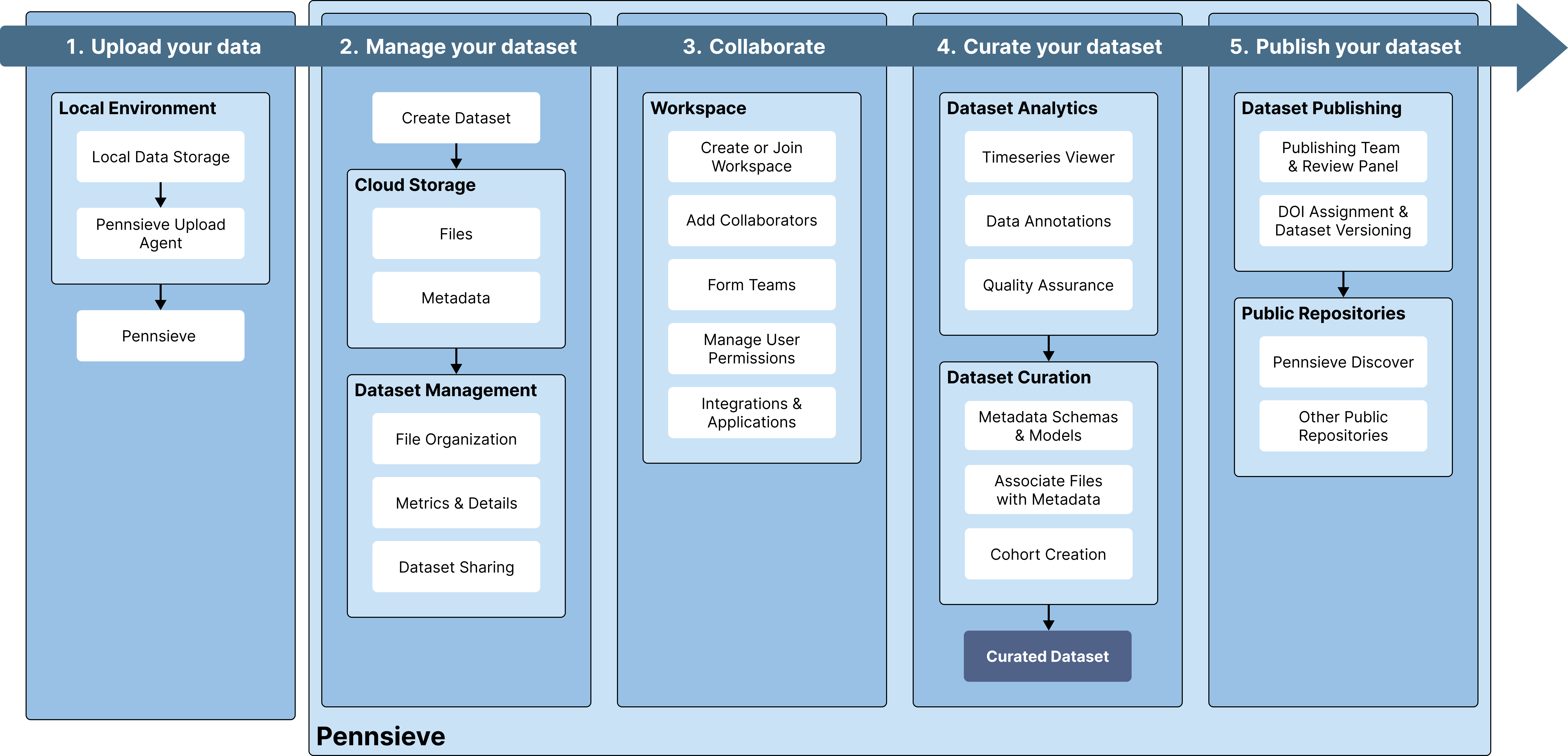}
\caption{Pennsieve's end-to-end data management workflow. The platform supports the full lifecycle of scientific data, from initial upload to final publication in public repositories.}
\label{fig:data_workflow}
\end{figure}

Pennsieve supports the life cycle of scientific data from initial upload through publication. The user workflow, detailed in Fig. \ref{fig:data_workflow}, encompasses several key functionalities: secure data upload, dataset management, collaboration within workspaces, dataset curation, and dataset publishing.
\begin{enumerate}
     \item Pennsieve supports file uploads through both its web application and programmatic methods. The upload process involves three steps: 1) generation of an upload manifest, 2) uploading the files, and 3) verification of the uploaded manifest. Large datasets should be uploaded programmatically to the platform and the Pennsieve Agent - an installable application for Windows, macOS, and Linux - facilitates this process. The agent manages manifest generation and file uploading to the platform. Several tools are available to interact with the agent, including a Python client, CLI tool, and GO library. The agent exposes a gRPC interface which makes it easy for users to develop integrations in other programming languages. Prior to uploading files, the manifest is synchronized with the cloud, enabling users to track expected data uploads over time and resume uploads in case of a client disconnection.  
    \item Data on Pennsieve are stored using AWS S3 cloud object storage and made available to users in the form of datasets, the primary components of Pennsieve. Datasets have a directory structure and actions such as renaming, moving, downloading, or deleting files and folders can be done through the web interface or programmatically via the Pennsieve API. Data file types are automatically analyzed and when applicable, converted into proprietary packages that enable direct interactions with the data on Pennsieve. Additional functionality is available for such files, and this derived data is stored alongside the original files. Metrics including total number of files, dataset size, and last updated date are available in an overview pane that provides a comprehensive summary of the dataset. Numerous dataset attributes can be specified: dataset name, subtitle, contributors, description, license, tags, and banner image. Users can keep datasets private or choose to share them with collaborators or entire workspaces on Pennsieve. 
    \item Workspaces are shared environments that users can create, or join, where datasets and tools can be collectively utilized. In these workspaces, users can share datasets, add collaborators, manage user permissions through role-based access controls, and organize into teams. This framework provides control over who can view, edit, and manage specific datasets. Discussion functionality integrated within timeseries and imaging viewers lets users collaborate alongside their visualized data. 
    \item The platform provides a set tools to associate complex metadata with datasets and files. Users can define a metadata schema for each dataset and create records linking to files. For example, a user could define a metadata schema of "Patients", "Hospital Visits", and "Samples", which are all linked and point to specific files. Or a different schema could be defined with "Animal", "Experiment", and "Trial" to capture a very different type of scientific dataset. This lets researchers define cohorts within their datasets, allowing for targeted studies and analyses. This capability is particularly useful for clinical and biomedical research, where grouping data by specific characteristics (e.g., patient demographics or treatment groups) can provide valuable insights. The outcome of this continuous curation process is high-quality datasets suitable for reproducible scientific research.
    \item The final stage of the workflow is publishing datasets to public repositories, making them accessible, citable, and reusable by the scientific community. The data publishing process is fully detailed in the following section. 
\end{enumerate}

\subsection{Data Publishing}
\begin{figure}[!htbp]
\centering
\includegraphics[width=0.8\textwidth]{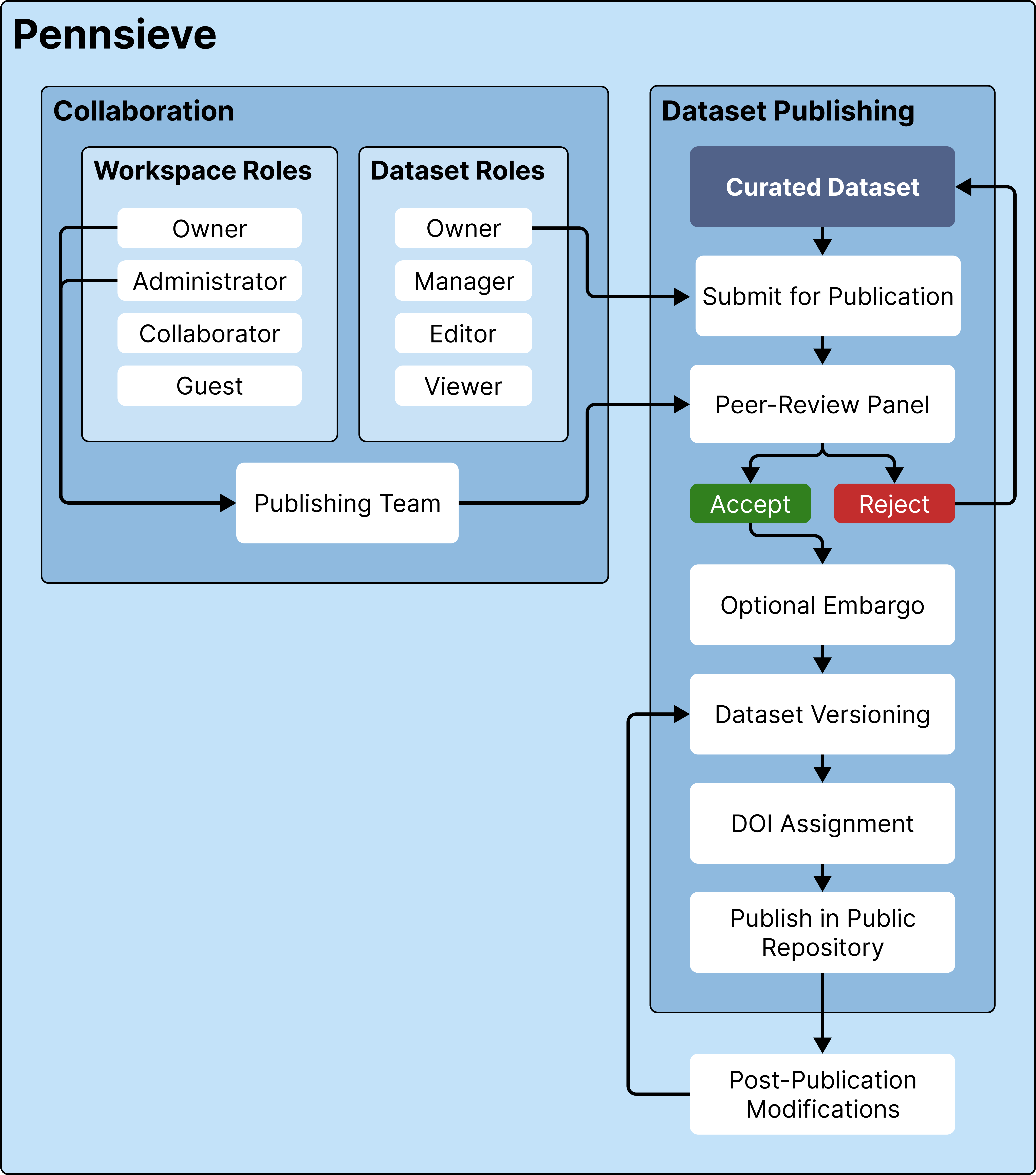}
\caption{Pennsieve's data publishing workflow. The progression from curated dataset to public repository includes peer review, version control, and DOI assignment.}
\label{fig:publishing_workflow}
\end{figure}

The data publishing process on Pennsieve (Fig. \ref{fig:publishing_workflow}) is how curated datasets get approved and finalized for distribution in public repositories. When a dataset reaches a publication-ready stage, the owner of the dataset can submit a request in their workspace for dataset peer-review. A publishing team, comprised of selected users with owner or administrator privileges, receives this request for evaluation. If the dataset is rejected, it can be revised for re-submission, while accepted datasets proceed in the publication process. A DOI is assigned to the dataset and it becomes version controlled. An optional embargo period of up to one year can be applied, giving authors control over when their dataset becomes publicly accessible. Finally, once all requisite steps are completed, the dataset is published in the chosen public repository. Any changes published after this point are controlled, with the modified dataset receiving a new version number and DOI. 

\subsection{Documentation Guides, Tutorials, and Recipes}
Multiple guides, tutorials, and runnable code snippets are available in the \href{https://docs.pennsieve.io/}{\textbf{\textit{Pennsieve Documentation Hub}}}. This includes instructions for how to set up accounts, workspaces, and manage datasets from upload through publication. \href{https://docs.pennsieve.io/recipes}{\textbf{\textit{Recipes}}} - manuals which break down source code into explainable chunks - detail common interactions with the platform. 

\subsection{Pennsieve API Reference}
Pennsieve has a well-documented and rich API. The \href{https://docs.pennsieve.io/reference/}{\textbf{\textit{API Reference}}} provides detailed information on the available API calls that allow for extensive interactions with the platform. 

\FloatBarrier
\subsection{Open-Source Code and Microservices}
Pennsieve is fully open-source and built with a microservices architecture. The \href{https://github.com/Pennsieve}{\textbf{\textit{Pennsieve codebase}}} is deployed to the AWS cloud in an Infrastructure as a Service (IaaS) model using Terraform and Jenkins. It is built as a collection of independent microservices that exchange information with each other and provide a uniform experience to the user. Here are the primary Pennsieve code repositories:
\begin{itemize}

    \item \href{https://github.com/Pennsieve/authentication-service}{\textbf{\textit{Authentication Service}}} - Manages user and data permissions using AWS Cognito (CIAM).
    \item \href{https://github.com/Pennsieve/pennsieve-app}{\textbf{\textit{Pennsieve App}}} and \href{https://github.com/Pennsieve/pennsieve-discover-app}{\textbf{\textit{Pennsieve Discover App}}} - Manage web applications and content visible to the users.
    \item \href{https://github.com/Pennsieve/pennsieve-api}{\textbf{\textit{Pennsieve API}}} and \href{https://github.com/Pennsieve/pennsieve-go-api}{\textbf{\textit{Pennsieve GO API}}} - Allow interaction with Pennsieve through its API.
    \item \href{https://github.com/Pennsieve/discover-service}{\textbf{\textit{Discover Service}}} - Manages the collection of public datasets.
    \item \href{https://github.com/Pennsieve/doi-service}{\textbf{\textit{DOI Service}}} - Creates, manages, and publishes DataCite DOIs.
    \item \href{https://github.com/Pennsieve/model-service}{\textbf{\textit{Model Service}}} and \href{https://github.com/Pennsieve/model-service-serverless}{\textbf{\textit{Serverless Model Service}}} - Provide a graph layer to the data (metadata and annotations).
    \item \href{https://github.com/Pennsieve/discover-publish}{\textbf{\textit{Discover Publish}}}, \href{https://github.com/Pennsieve/publishing-service}{\textbf{\textit{Publishing Service}}}, and \href{https://github.com/Pennsieve/datasets-service}{\textbf{\textit{Datasets Service}}} - Handle publishing and managing datasets.
    \item \href{https://github.com/Pennsieve/discover-release}{\textbf{\textit{Discover Release}}} - Manages embargoed and private datasets.
    \item \href{https://github.com/Pennsieve/pennsieve-agent}{\textbf{\textit{Pennsieve Agent}}}, \href{https://github.com/Pennsieve/pennsieve-agent-python}{\textbf{\textit{Pennsieve Agent Python}}}, and \href{https://github.com/Pennsieve/pennsieve-agent-javascript}{\textbf{\textit{Pennsieve Agent JavaScript}}} - Interfaces in GO, Python, and JavaScript that interact with the platform and allow users to programmatically upload and download files.
    \item \href{https://github.com/Pennsieve/workflow-manager}{\textbf{\textit{Workflow Manager}}}, \href{https://github.com/Pennsieve/etl-nextflow}{\textbf{\textit{ETL Nextflow}}}, and \href{https://github.com/Pennsieve/app-deploy-service}{\textbf{\textit{App Deploy Service}}} - Create, orchestrate, and manage applications and workflows on Pennsieve.
    \item \href{https://github.com/Pennsieve/rehydration-service}{\textbf{\textit{Rehydration Service}}} - Retrieves datasets or files from any of the previously released versions and prepares them for download.
    \item \href{https://github.com/Pennsieve/integration-service}{\textbf{\textit{Integration Service}}} - Creates webhooks and manages integrations and notifications.
    \item \href{https://github.com/Pennsieve/timeseries-processor}{\textbf{\textit{Timeseries Processor}}} and \href{https://github.com/Pennsieve/pennsieve-streaming}{\textbf{\textit{Pennsieve Streaming Service}}} - Ingest data to Pennsieve and stream, download, and display data.
    \item \href{https://github.com/Pennsieve/app-deploy}{\textbf{\textit{App Deploy Service}}} - Sets up infrastructure in AWS.
    \item \href{https://github.com/Pennsieve/upload-service}{\textbf{\textit{Upload Service}}} - Uploads large files to Pennsieve. 
    
\end{itemize}
\FloatBarrier

\subsection{Data Security}
Built on AWS S3, Pennsieve utilizes cloud infrastructure to provide high scalability, security, and availability for file storage. Individual files up to 5TB in size are supported and an essentially unlimited number of files can be handled. AWS S3 storage provides 99.999999999\% (11 nines) annual durability protection from loss or corruption and 99.99\% (4 nines) availability. All uploaded files undergo checksum testing and are secured at rest using server-side encryption (SSE). Data transfers on Pennsieve are secured with Secure Sockets Layer (SSL) encryption to protect it from interception and unauthorized access. Pennsieve supports encryption keys on a per-workspace basis and distributed data storage to further compartmentalize access if necessary for security or compliance reasons. These systems safely and reliably persist data at scale.

\subsection{Governance and Sustainability}
Pennsieve's governance structure has several mechanisms that ensure its sustainability, compliance with standards, and responsiveness to user needs.

\subsubsection{Advisory Boards}
Three advisory boards guide Pennsieve:
\begin{enumerate}
    \item Clinical Advisory Board: Provides guidance on data standards, clinical workflows, and how to enhance the platform's impact on patient care. This advisory board is currently being established in step with the platform's increasing application in clinical research. 
    \item Technical Advisory Board: Advises on implementing technical standards, developing the platform's roadmap, and integrating with other data science efforts.
    \item Team Blue: Prioritizes non-critical feature requests and provides user feedback that aligns Pennsieve's development with user needs.
\end{enumerate}

The technical and clinical advisory boards meet biannually to steer Pennsieve's strategic direction and include representatives from academia, non-profit organizations, and industry partners. They are responsible for reviewing and approving policies related to the platform, including data submission criteria and standards for dataset inclusion. Team Blue is a user-based advisory panel that convenes every two months and is comprised of at least four external and two Penn-based investigators actively using Pennsieve.

\subsubsection{Compliance and Certifications}
Pennsieve meets all required and recommended aspects of FAIR data sharing and is in the process of obtaining the CoreTrustSeal certification. Additionally, Pennsieve is GDPR compliant and is developed with HIPAA-certification in mind. Currently, it does not claim HIPAA compliance and does not accept PHI data. The Pennsieve team is working with the University of Pennsylvania towards HIPAA compliance attestation.  

\subsubsection{Sustainability and Migration Strategies}
Pennsieve recognizes that open access to research data is crucial, but maintaining and providing this data entails significant costs, especially given the increasing size of research datasets. To keep data open access while balancing operational costs, a sustainability strategy is in place with four primary directives: cost reduction while maintaining scalability, efficient maintenance, smart data storage, and value-driven user support. To achieve these, Pennsieve employs a mix of Docker-based services and serverless architectures that can rapidly scale up, but remain cost-free when inactive. The platform does continuous integration and automated deployments using industry standard technologies such as Terraform, Puppet, Jenkins, Nexus, and GitHub Actions. These maintain operational integrity while minimizing technical support costs.

For data storage, Pennsieve implements cloud-storage with automated versioning of files, tiered storage (active, archive, and deep archive), and requester-pays access for very large datasets. It accepts datasets under 25 gigabytes for publication free of charge, guaranteeing their availability for 10 years. Larger datasets, particularly those exceeding 1 terabyte, require justification and may need external funding through a data sharing plan.

Pennsieve's long-term financial sustainability is supported in several ways. Multiple projects and repositories share the Pennsieve infrastructure, effectively distributing operational costs. The Penn Institute for Biomedical Informatics and the Penn CNT, in addition to several federal funding sources, provide long-term financial support for operating costs. The platform also has mechanisms in place to charge for dataset submissions on an individual basis.

Pennsieve has disaster recovery procedures to ensure data integrity and availability. These include daily database backups with multiple backups kept on a rolling basis, event-logs for full state recovery, file versioning through AWS with 30-day undelete support, and deployment across multiple AWS availability zones to mitigate risks of sporadic failures or downtime. For long-term data management, Pennsieve implements a data sunsetting policy where data is transferred to lower-tier storage after a period of inactivity. Archived data is significantly cheaper to store, with restoration costs incurred when accessed. Pennsieve is also developing unarchiving mechanisms at requester expense to keep all data available on the platform beyond the initial funding horizon.

In the unlikely event that the University of Pennsylvania can no longer support Pennsieve, migration strategies are in place. These cover scenarios ranging from a new maintainer without Pennsieve capabilities to full platform migration. Even without the platform infrastructure, published data can remain available to researchers and compliant with the FAIR principles. Pennsieve's data publishing strategy results in self-describing datasets on AWS S3, with all data and metadata serialized into a set of files during the publication process. Data accessibility can be maintained by simply redirecting DOIs to the JSON-structured manifest files for each published dataset on S3.

\clearpage
\section{How can I contribute?}
Pennsieve is committed to high standards of data quality and encourages members of the scientific community to contribute their data to the platform.

\subsection{Researcher}
Pennsieve serves researchers as a platform for publishing, sharing, and annotating data. Documentation on setting up an account can be found in the \href{https://docs.pennsieve.io/docs/getting-started}{\textbf{\textit{Getting Started with Pennsieve}}} tutorial. Researchers may upload their files to the platform using the Pennsieve Agent (see \href{https://docs.pennsieve.io/docs/uploading-files-using-the-pennsieve-agent}{\textbf{\textit{tutorial}}}). This is the recommended method for large or complex datasets. The Pennsieve Agent is a local, lightweight client that controls and verifies data transfer to cloud storage in Pennsieve. It allows users to prepare files for upload in a package and monitors the upload status. Currently, Python, JavaScript, and CLI clients are supported on all major platforms (Windows/macOS/Linux).

Researchers can use several features on Pennsieve to create and contribute datasets. \href{https://docs.pennsieve.io/docs/pennsieve-web-application}{\textbf{\textit{Collaborative workspaces}}} are shared environments where datasets and tools can be utilized collectively. Researchers can manage user permissions through role-based access controls and organize their members into teams. \href{https://docs.pennsieve.io/docs/data-management-feature-set}{\textbf{\textit{Dataset curation}}} tools allow direct data and metadata management. \href{https://docs.pennsieve.io/docs/overview}{\textbf{\textit{Publication pipelines}}} enable publishing to public repositories, making datasets accessible, citable, and reusable by the scientific community.

\subsection{Data Analyst}
Pennsieve streamlines downloading and managing data by providing a standard, open API. Requests to and responses from the server can be easily tested in 20 different programming languages; please refer to the \href{https://docs.pennsieve.io/reference/}{\textbf{\textit{Pennsieve API Reference}}} for more details.

Downloading publicly available files is straightforward using the Pennsieve Agent. Examples for querying and obtaining data from Pennsieve can be found in the \href{https://github.com/nih-sparc/sparc.client/blob/main/docs/tutorial.ipynb}{\textbf{\textit{NIH SPARC Python Client Tutorial}}}. Documentation on its functions is available in the \href{https://nih-sparc.github.io/sparc.client/sparc.client.services.html#module-sparc.client.services.pennsieve}{\textbf{\textit{Package Reference}}}.

Data analysts are encouraged to share their analytic workflows. By contributing scripts, pipelines, and methodologies to the Pennsieve community, data analysts support reproducible research.

\subsection{Developer}
Pennsieve's open-source codebase allows developers to contribute by fixing bugs, adding features, or improving documentation. Additionally, using the \href{https://github.com/Pennsieve/template-serverless-service}{\textbf{\textit{serverless service template}}}, developers can create microservices that extend Pennsieve's capabilities, offering new tools to the community. The Pennsieve API can be used to add new endpoints, improve existing ones, or add integrations with other tools and platforms.

\subsection{Data Annotator}
Pennsieve provides user-friendly tools for annotating \href{https://docs.pennsieve.io/docs/timeseries-viewer}{\textbf{\textit{timeseries}}} and \href{https://docs.pennsieve.io/docs/generic-data-viewers}{\textbf{\textit{clinical imaging}}} data. These annotations can be layered over the data, adding additional context and insights. Annotators can create relationships between metadata models and link records to specific files, creating a structured, searchable dataset. A graph metadata viewer makes it easier to understand complex datasets by visualizing the relationships between data and metadata.

\subsection{User}
Users have free access to data through \href{https://discover.pennsieve.io/}{\textbf{\textit{Pennsieve Discover}}}. They can browse and download publicly available datasets from a range of scientific fields. Each dataset includes the metadata, annotations, and related documents necessary for its reuse. Search tools help users find datasets relevant to their research interests, with options to filter results by tags and dataset status. Additionally, they can engage with the community by participating in \href{https://docs.pennsieve.io/discuss}{\textbf{\textit{discussions}}} and initiating collaborations with dataset authors. 

\clearpage
\section{Datasets on Pennsieve}

\subsection{Dataset Diversity}
\begin{figure}[!hbtp]
    \centering
    \colorbox{white}{%
    \begin{minipage}{\textwidth}
    \begin{subfigure}[b]{0.48\textwidth}
        \centering
        \includegraphics[width=\textwidth]{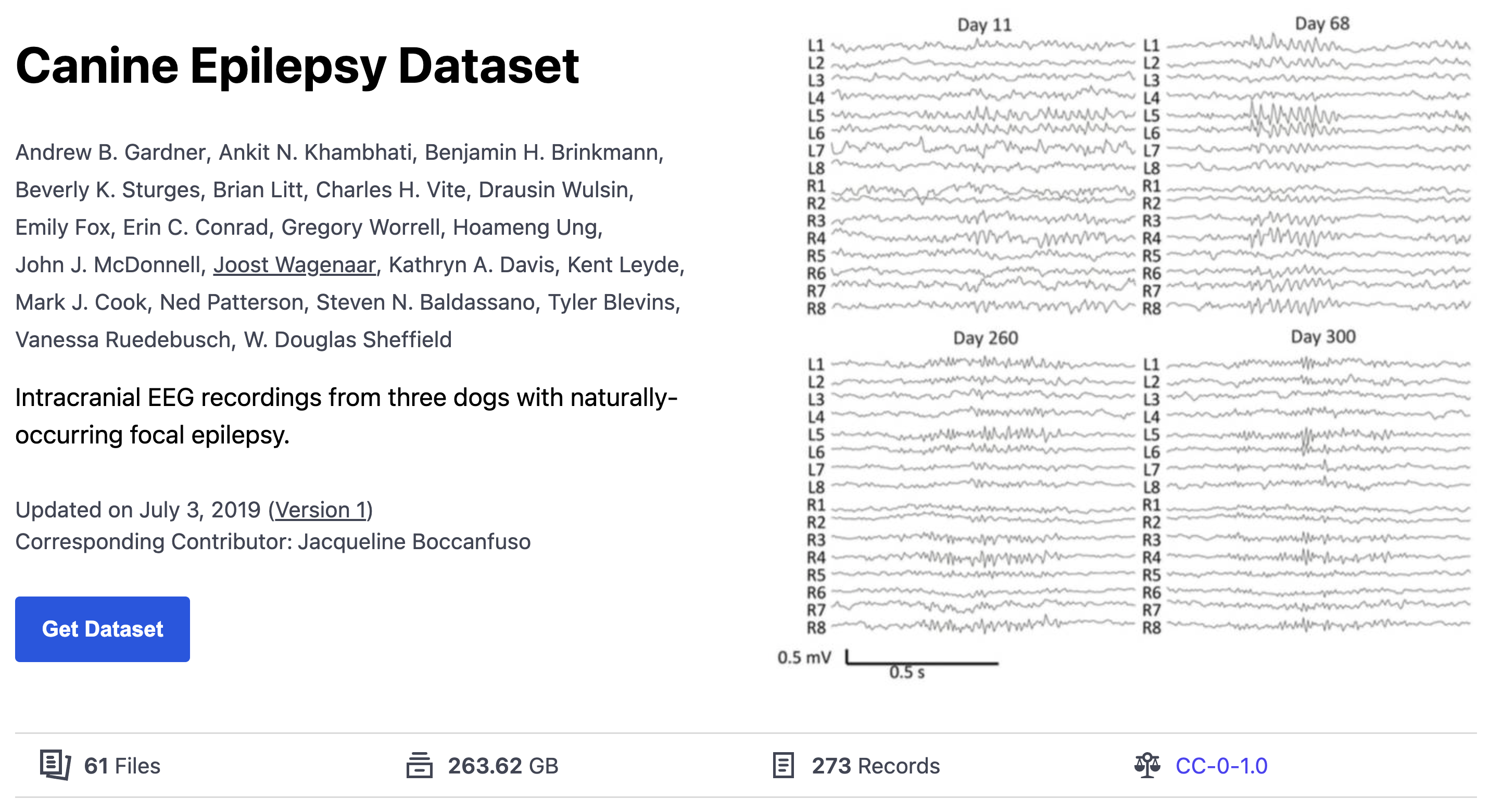}
        \caption{Electrophysiology \cite{Gardner2019}.}
        \label{fig:image1}
    \end{subfigure}
    \hfill
    \begin{subfigure}[b]{0.48\textwidth}
        \centering
        \includegraphics[width=\textwidth]{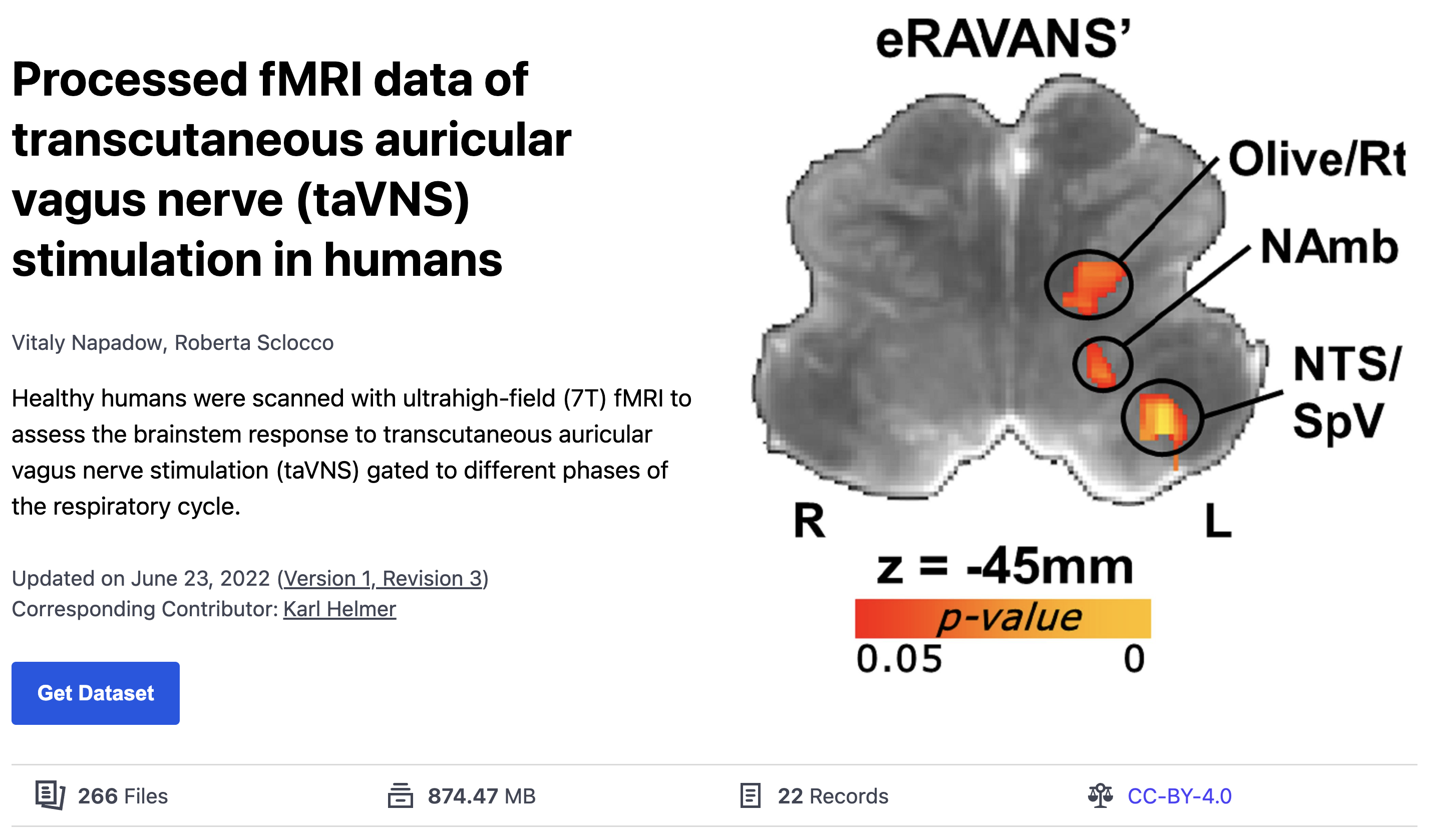}
        \caption{Clinical imaging \cite{Napadow2020}.}
        \label{fig:image2}
    \end{subfigure}
    \vfill
    \vspace{0.5cm}
    \begin{subfigure}[b]{0.48\textwidth}
        \centering
        \includegraphics[width=\textwidth]{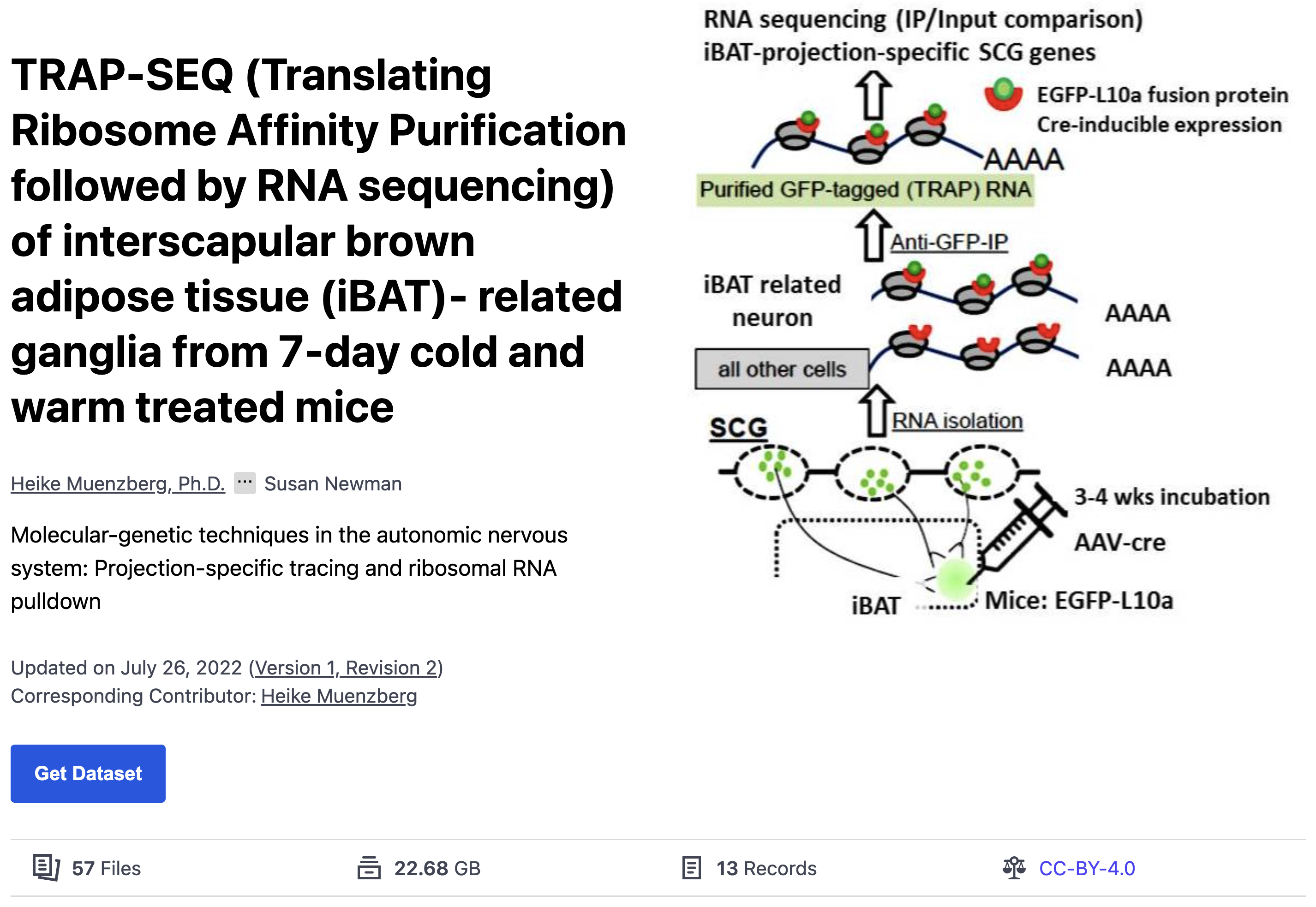}
        \caption{Genetic information \cite{Muenzberg2021}.}
        \label{fig:image3}
    \end{subfigure}
    \hfill
    \begin{subfigure}[b]{0.48\textwidth}
        \centering
        \includegraphics[width=\textwidth]{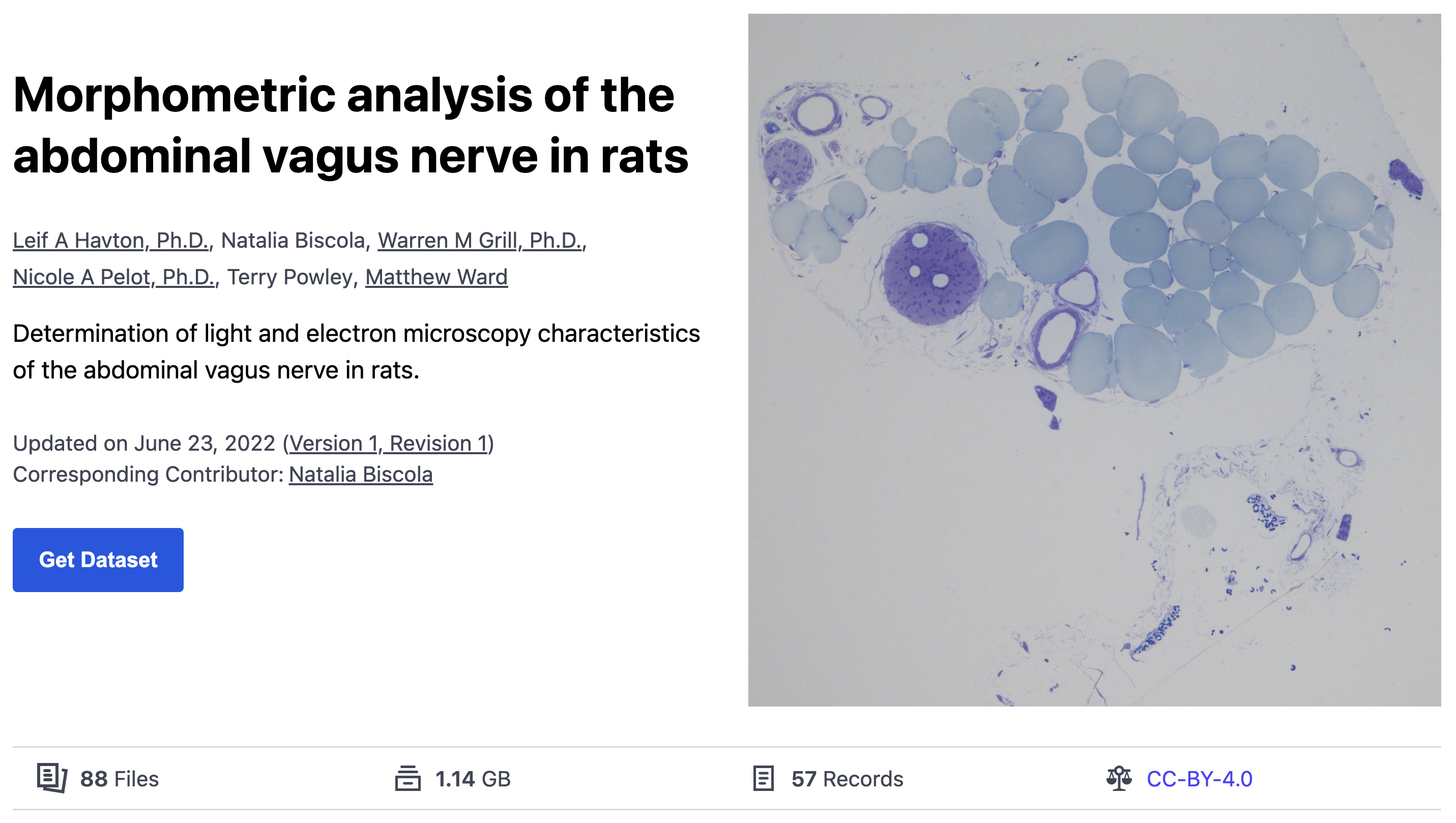}
        \caption{Microscopy \cite{Havton2020}.}
        \label{fig:image4}
    \end{subfigure}
    \vfill
    \vspace{0.5cm}
    \begin{subfigure}[b]{0.48\textwidth}
        \centering
        \includegraphics[width=\textwidth]{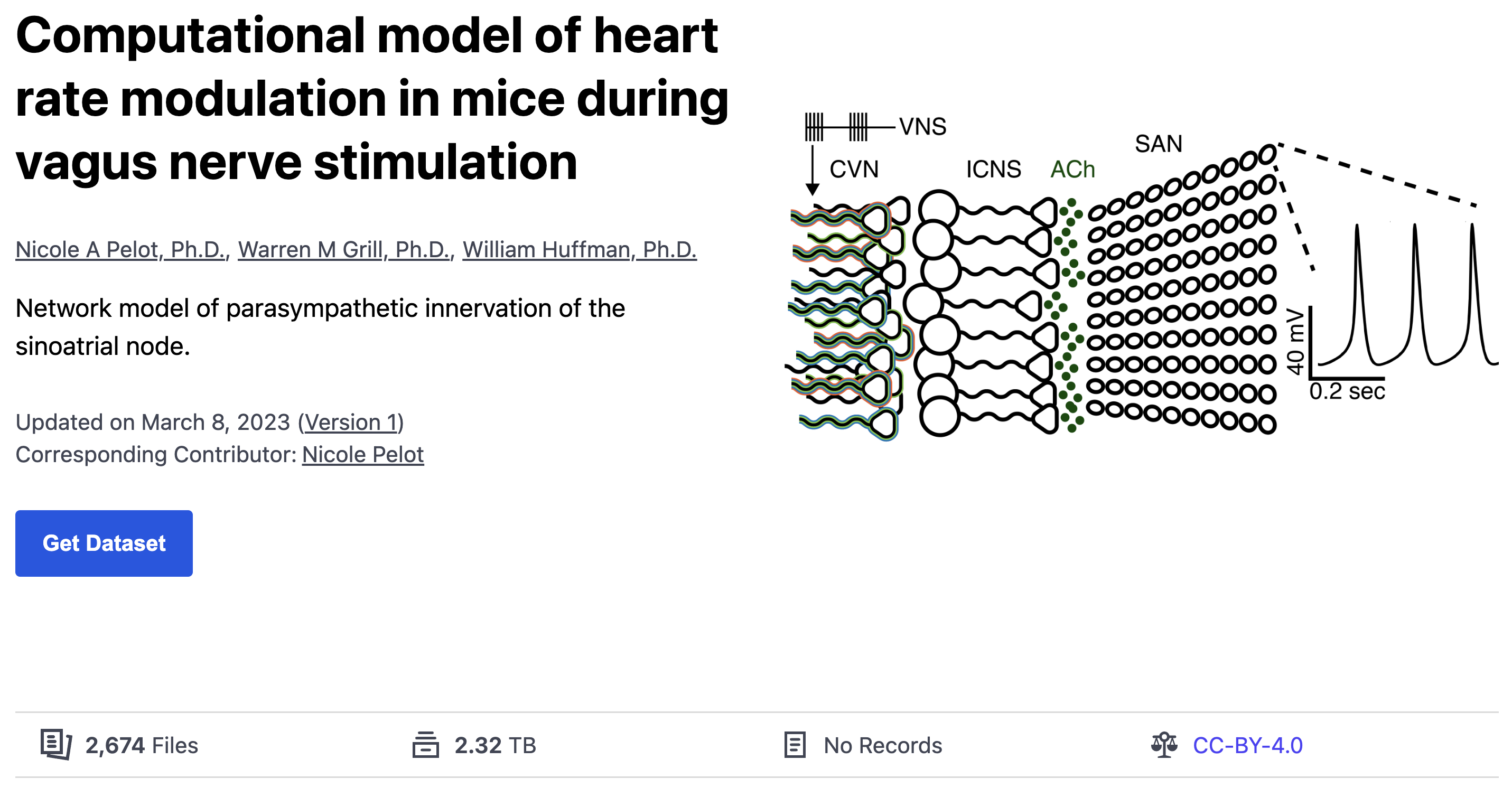}
        \caption{Computational models \cite{Pelot2023}.}
        \label{fig:image5}
    \end{subfigure}
    \hfill
    \begin{subfigure}[b]{0.48\textwidth}
        \centering
        \includegraphics[width=\textwidth]{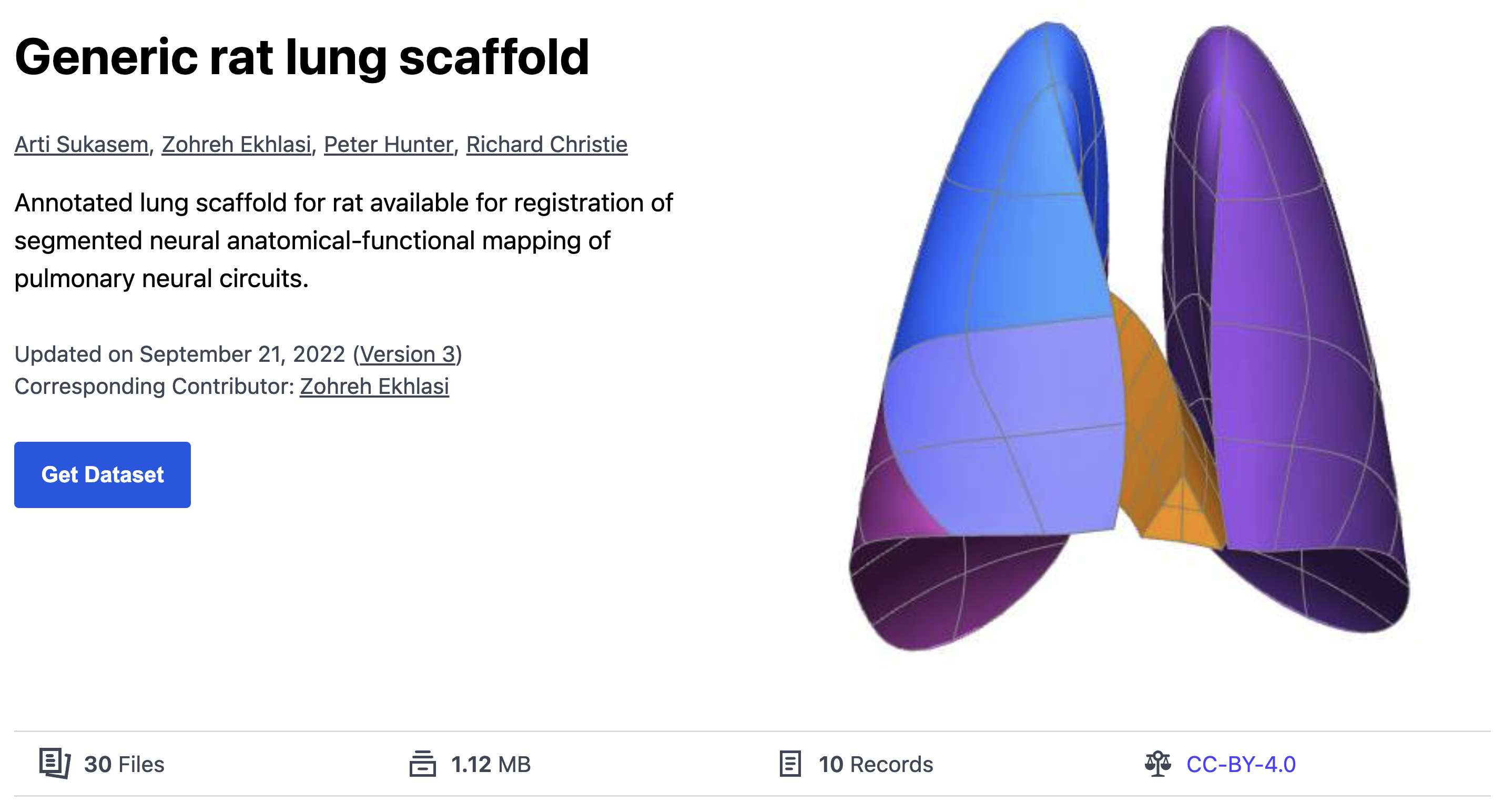}
        \caption{Anatomical models \cite{Sukasem2022}.}
        \label{fig:image6}
    \end{subfigure}
    \end{minipage}%
    }
    \caption{Selection of diverse neuroscience datasets published on Pennsieve.}
    \label{fig:example_datasets}
\end{figure}

Pennsieve hosts over 350 publicly available datasets. These datasets span many neuroscience topics: vagus nerve stimulation (VNS), neural circuits, synaptic activity, electrophysiology, neurotransmitter systems, neural devices, neural imaging, computational models, gene expression, neural development, plasticity, and neural injury and repair. They provide a comprehensive resource for the neuroscience research community across multiple modalities including EEG, MEG, MRI, microscopy images, gene data, 3D models, and videos. Highlights of some datasets hosted on Pennsieve are presented in Fig. \ref{fig:example_datasets}.

\subsection{Data Viewing and Annotations}
\begin{figure}[!hbtp]
    \centering
    \colorbox{white}{%
    \begin{minipage}{0.9\textwidth}
    \begin{subfigure}[b]{\textwidth}
        \centering
        \includegraphics[width=\textwidth]{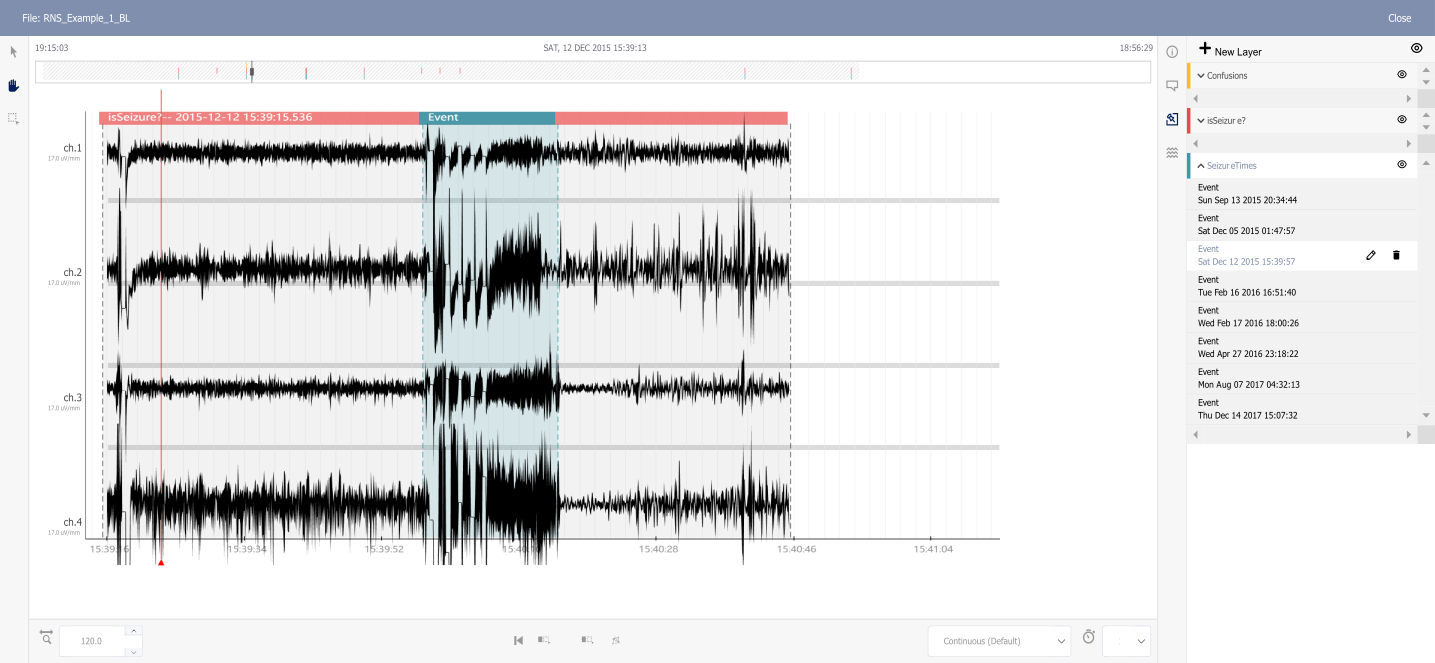}
        \caption{Timeseries viewer with annotations.}
        \label{fig:image7}
    \end{subfigure}
    \vfill
    \vspace{0.5cm}
    \begin{subfigure}[b]{\textwidth}
        \centering
        \includegraphics[width=\textwidth]{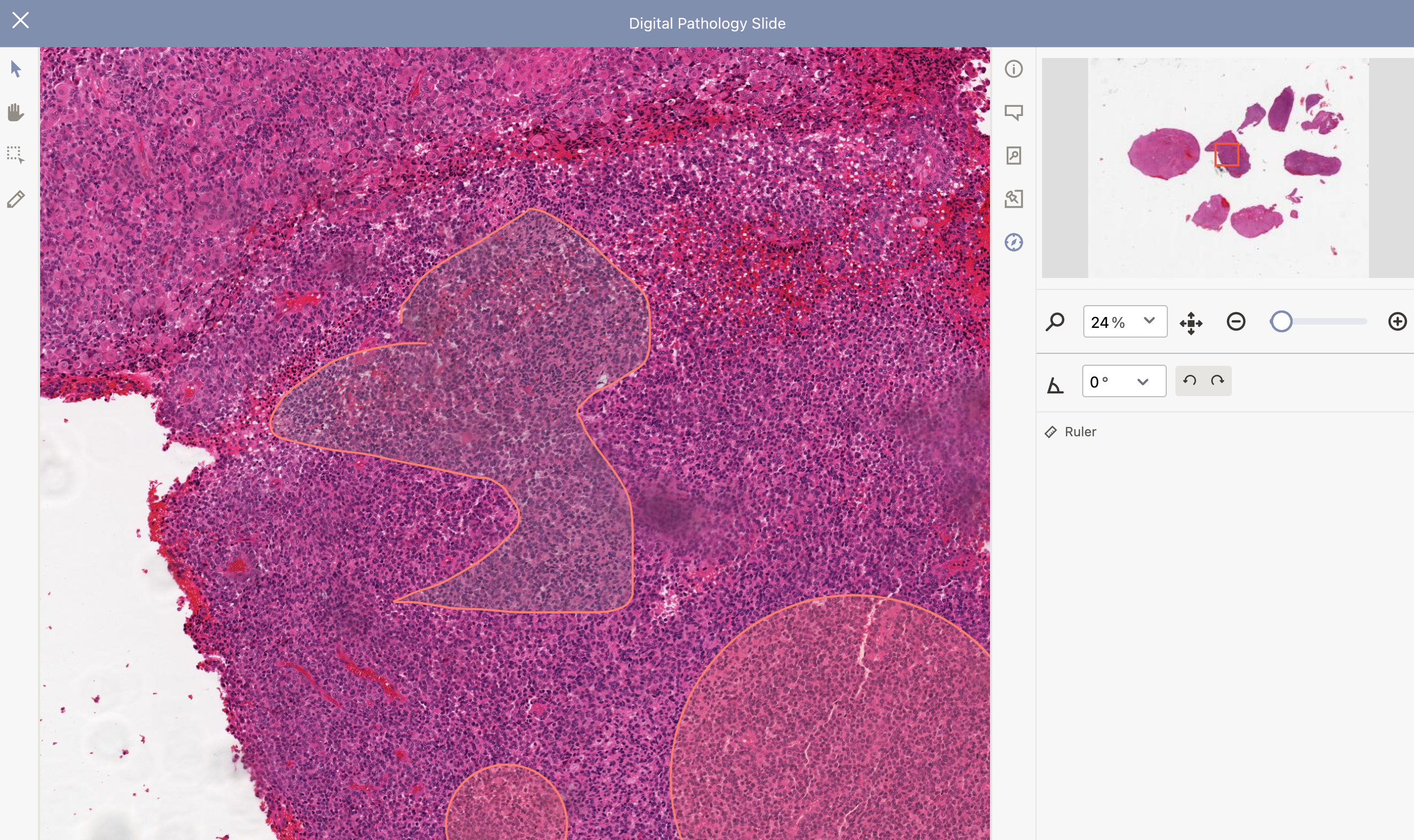}
        \caption{Imaging viewer with annotations.}
        \label{fig:image8}
    \end{subfigure}
    \end{minipage}%
    }
    \caption{Integrated data viewers with annotation tools on Pennsieve.}
    \label{fig:viewers}
\end{figure}

In scientific data analysis, annotations are crucial to the data themselves. Recognizing this, Pennsieve has developed viewers for various data modalities that allow users to view both data and annotations directly on the platform. These data viewers feature integrated tools for adding new annotations to datasets as overlaying layers. Fig. \ref{fig:viewers} provides examples of these viewers and annotation tools. More details about the viewers can be found in the \href{https://docs.pennsieve.io/docs/timeseries-viewer}{\textbf{\textit{documentation}}}. 

\FloatBarrier
\subsection{Dataset Structure}
\begin{figure}[!hbtp]
    \centering
    \includegraphics[width=0.65\textwidth]{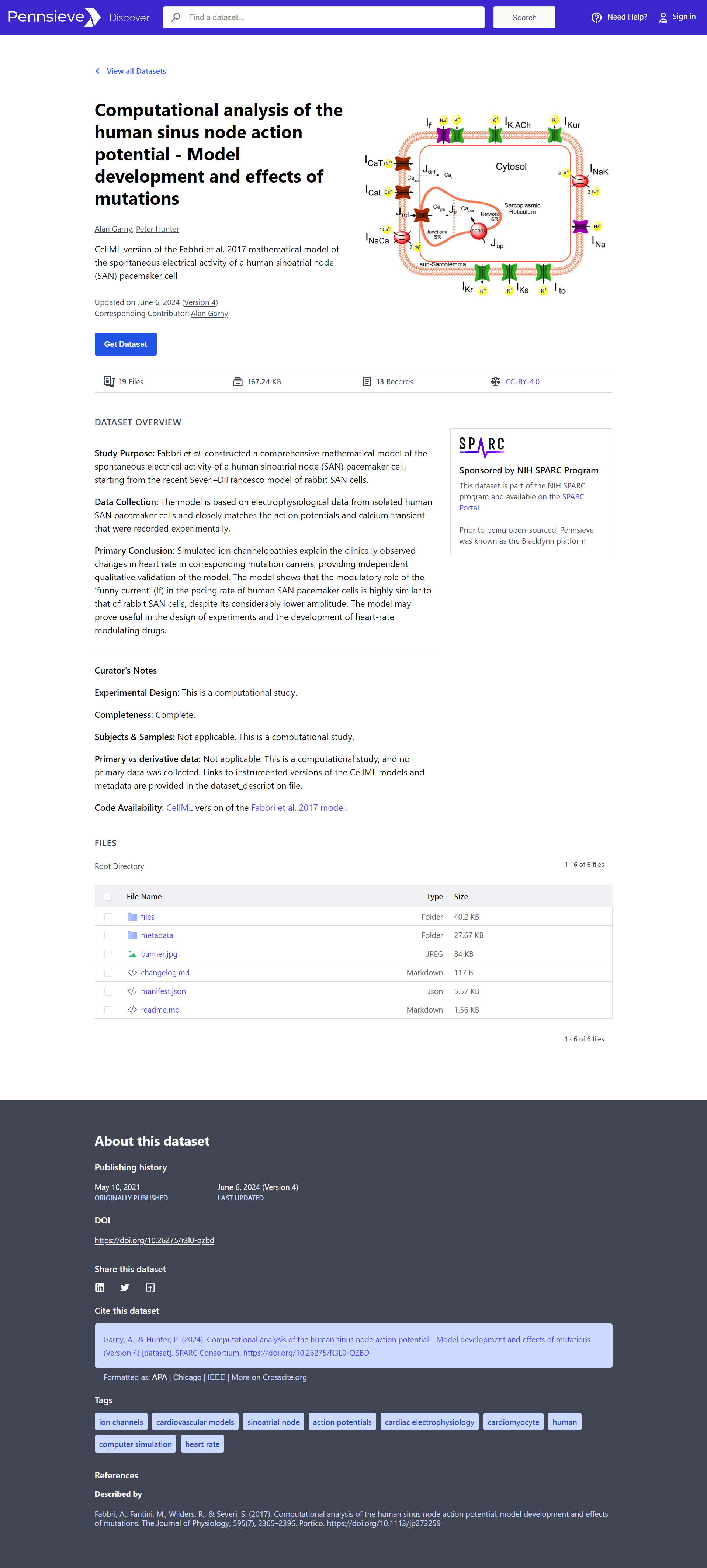}
    \caption{Example of a published dataset page on Pennsieve Discover.}
    \label{fig:dataset_page_structure}
\end{figure}

Pennsieve Discover organizes each dataset page with a user-centric design, ensuring researchers can easily access its content and understand the scope and relevance to their own work. An example dataset page on Pennsieve Discover is shown in Fig. \ref{fig:dataset_page_structure}. The entire dataset, or individual files from it, can be downloaded directly from its page. The key sections include:

\begin{itemize}
    \item Header
    \begin{itemize}
        \item Title: Name of the dataset.
        \item Contributors: Individuals who contributed to the dataset, along with their affiliations and roles.
        \item Description: A brief summary of the dataset.
    \end{itemize}
    \item Metrics
    \begin{itemize}
        \item Number of files, dataset size, number of metadata records, and the license.
    \end{itemize}
    \item Dataset Overview
    \begin{itemize}
        \item Study Purpose: Describes the scientific objective and background of the study.
        \item Data Collection: Details the methods and protocols used to gather the data.
        \item Primary Conclusion: Summarizes the main findings and significance of the dataset.
    \end{itemize}
    \item Curator's Notes
    \begin{itemize}
        \item Experimental Design: Specifies whether the study is experimental, observational, or computational.
        \item Completeness: Indicates whether the dataset is complete or ongoing.
        \item Subjects \& Samples: Provides details on the subjects or samples used in the study, if applicable.
        \item Primary vs Derivative Data: Clarifies whether the dataset contains primary data, derivative data, or both.
        \item Code Availability: Links to any code or models associated with the dataset.
    \end{itemize}
    \item Files
    \begin{itemize}
        \item Displays a directory structure of the dataset, including data files, metadata, and any additional documents such as readme files or changelogs.
    \end{itemize}
    \item About
    \begin{itemize}
        \item Publishing History: Dates of original publication and last modification, also specifying the current version of the dataset.
        \item Citation \& Sharing: The DOI for the dataset, the appropriate citation in several formats, and shortcuts to share the dataset on social platforms.
        \item Tags: Relevant keywords and tags associated with the dataset for easier discovery and categorization.
        \item References: Citations of literature that utilize this dataset.
    \end{itemize}
\end{itemize}
\FloatBarrier

\begin{landscape}
\subsection{Pennsieve's Current Data Repository}

\FloatBarrier
The following table contains public datasets available on the platform as of July 3, 2024. 

\renewcommand{\arraystretch}{1.2}
\begin{longtable}{p{0.18\linewidth} p{0.18\linewidth} p{0.46\linewidth} p{0.08\linewidth}}
\toprule
    DOI & Owner & Title & Year \\
\midrule
\endhead
\href{http://doi.org/10.26275/b0lj-xsto}{10.26275/b0lj-xsto} & Adeodu, O. & Simulation of the short term impact of atrial fibrillation on hemodynamic variables & 2024 \\
\href{http://doi.org/10.26275/qach-onkl}{10.26275/qach-onkl} & Keast, J. & Selective recording of physiologically evoked neural activity in a mixed autonomic nerve using a minimally invasive array & 2024 \\
\href{http://doi.org/10.26275/yues-hayo}{10.26275/yues-hayo} & Garny, A. & Fabbri-based composite SAN model & 2024 \\
\href{http://doi.org/10.26275/nbex-w34w}{10.26275/nbex-w34w} & Garny, A. & Computational analysis of the human sinus node action potential - Model development and effects of mutations & 2024 \\
\href{http://doi.org/10.26275/mgyp-rlmr}{10.26275/mgyp-rlmr} & Pelot, N. & Excitation properties of computational models of unmyelinated peripheral axons & 2024 \\
\href{http://doi.org/10.26275/yssz-iprs}{10.26275/yssz-iprs} & Lin, M. & Generic human stomach scaffold & 2024 \\
\href{http://doi.org/10.26275/s3l0-t915}{10.26275/s3l0-t915} & Lin, M. & A 3D human whole-body model with integrated organs vasculature musculoskeletal and nervous systems for mapping nerves & 2024 \\
\href{http://doi.org/10.26275/t2ve-vu9o}{10.26275/t2ve-vu9o} & Aguinaga, D. & Targeting bladder function with network-specific epidural stimulation after chronic spinal cord injury & 2024 \\
\href{http://doi.org/10.26275/mu8z-s80h}{10.26275/mu8z-s80h} & Arendt-Tranholm, A. & Long read sequencing of human dorsal root ganglia & 2024 \\
\href{http://doi.org/10.26275/rdr5-x5bo}{10.26275/rdr5-x5bo} & Ardell, J. & Transcriptomic and neurochemical analysis of the stellate ganglia in mice highlights sex differences & 2024 \\
\href{http://doi.org/10.26275/odx3-c5cv}{10.26275/odx3-c5cv} & Stebbing, M. & Quantification of enteric ganglia in the three main regions of the rat stomach & 2024 \\
\href{http://doi.org/10.26275/3enb-ctj6}{10.26275/3enb-ctj6} & Jaffey, D. & Vagal preganglionic axons arborize in myenteric plexus into two patterns generating nitrergic and non-nitrergic postganglionic motor units & 2024 \\
\href{http://doi.org/10.26275/hipp-yfnn}{10.26275/hipp-yfnn} & Wang, L. & Vasculature in mouse colon and relationship with enteric nervous system glia and macrophages & 2024 \\
\href{http://doi.org/10.26275/xw1e-oea3}{10.26275/xw1e-oea3} & Keast, J. & Light microscopic analysis of synaptic input to neurons in the rat major pelvic ganglion & 2024 \\
\href{http://doi.org/10.26275/pc5x-w5l3}{10.26275/pc5x-w5l3} & Wang, L. & Transduction of systemically administered adeno-associated virus in the colonic enteric nervous system of adult mice & 2024 \\
\href{http://doi.org/10.26275/1ito-vvqo}{10.26275/1ito-vvqo} & Nguyen, D. & Substance P-immunoreactive axons in the antrum-pylorus-duodenum of mice & 2024 \\
\href{http://doi.org/10.26275/cheq-otb2}{10.26275/cheq-otb2} & Wang, L. & Intravenously injected AAV9 transduced interstitial cells of Cajal in mouse colon & 2024 \\
\href{http://doi.org/10.26275/cljd-yjqy}{10.26275/cljd-yjqy} & Nguyen, D. & Calcitonin gene-related peptide - immunoreactive (CGRP-IR) axon innervation of mouse stomach & 2024 \\
\href{http://doi.org/10.26275/zphj-6qnb}{10.26275/zphj-6qnb} & Taylor-Clark, T. & Mapping of vagal sensory nerve populations and their brainstem projections in mice & 2024 \\
\href{http://doi.org/10.26275/aw2z-a49z}{10.26275/aw2z-a49z} & Armitage, O. & Influence of vagus nerve stimulation on vagal and cardiac activity in freely moving pigs & 2024 \\
\href{http://doi.org/10.26275/ixhd-0pfa}{10.26275/ixhd-0pfa} & Athavale, O. & A mathematical model for simulating the neural regulation of phasic contractions and slow waves in the distal stomach & 2024 \\
\href{http://doi.org/10.26275/2njc-khrp}{10.26275/2njc-khrp} & Venkataraman, A. & Effects of manipulating epithelial cell-nerve signaling on anxiety-like behavior in mice & 2024 \\
\href{http://doi.org/10.26275/2xys-dleh}{10.26275/2xys-dleh} & Pelot, N. & Electromyogram recordings in mice during vagus nerve stimulation & 2024 \\
\href{http://doi.org/10.26275/eauv-gxwl}{10.26275/eauv-gxwl} & Pelot, N. & Electrocardiogram recordings in mice during vagus nerve stimulation & 2024 \\
\href{http://doi.org/10.26275/q2w0-keol}{10.26275/q2w0-keol} & Williams, A. & Two-Color Diffuse In Vivo Flow Cytometer & 2024 \\
\href{http://doi.org/10.26275/86ve-meck}{10.26275/86ve-meck} & Bendowski, K. & Distribution and morphology of calcitonin gene-related peptide (CGRP) innervation in flat mounts of whole rat atria and ventricles & 2024 \\
\href{http://doi.org/10.26275/lcqj-dvpi}{10.26275/lcqj-dvpi} & Ardell, J. & Sex differences in sympathetic gene expression and cardiac neurochemistry in Wistar Kyoto rats & 2024 \\
\href{http://doi.org/10.26275/gud3-alhy}{10.26275/gud3-alhy} & Stebbing, M. & Quantitation of nodose neurons labelled by retrograde tracing from selective gastric mucosal and gastric muscle injection & 2024 \\
\href{http://doi.org/10.26275/n12j-ubmj}{10.26275/n12j-ubmj} & Stebbing, M. & RNAseq analysis of  nodose neurons projecting to specific regions and tissue layers within the rat stomach & 2024 \\
\href{http://doi.org/10.26275/79vx-xnat}{10.26275/79vx-xnat} & Stebbing, M. & Quantitation of nodose neurons labelled by retrograde tracing from selective injections into the gastric mucosa or muscle in the antrum and fundus & 2024 \\
\href{http://doi.org/10.26275/yv7l-9fst}{10.26275/yv7l-9fst} & Howard, M. & Synaptic components function and modulation characterized by GCaMP6f calcium imaging in cholinergic myenteric ganglion neurons & 2024 \\
\href{http://doi.org/10.26275/tlgo-dfke}{10.26275/tlgo-dfke} & Stebbing, M. & Quantification of thickness of the gastric muscle in the rat stomach & 2024 \\
\href{http://doi.org/10.26275/mhji-cn6n}{10.26275/mhji-cn6n} & Stebbing, M. & Morphologies dimensions and targets of gastric nitric oxide synthase neurons in the rat stomach & 2024 \\
\href{http://doi.org/10.26275/3d2g-d3xd}{10.26275/3d2g-d3xd} & Iavarone, E. & ASCENT Tutorial & 2023 \\
\href{http://doi.org/10.26275/0jz3-zrlo}{10.26275/0jz3-zrlo} & Iavarone, E. & ASCENT Guided Mode Demo & 2023 \\
\href{http://doi.org/10.26275/yanw-4h7n}{10.26275/yanw-4h7n} & Iavarone, E. & ASCENT Base & 2023 \\
\href{http://doi.org/10.26275/qdv4-0vx5}{10.26275/qdv4-0vx5} & Bizanti, A. & Chronic intermittent hypoxia remodels catecholaminergic innervation in mouse atria & 2023 \\
\href{http://doi.org/10.26275/pmnr-f6br}{10.26275/pmnr-f6br} & Stacey, W. & Michigan High Res EEG - UMHS0018 & 2023 \\
\href{http://doi.org/10.26275/jkbj-f3bn}{10.26275/jkbj-f3bn} & Stacey, W. & Michigan High Res EEG - UMHS 0028 & 2023 \\
\href{http://doi.org/10.26275/s3jn-eo59}{10.26275/s3jn-eo59} & Stacey, W. & Michigan High Res EEG - UMHS 0023 & 2023 \\
\href{http://doi.org/10.26275/73j6-8nej}{10.26275/73j6-8nej} & Stacey, W. & Michigan High Res EEG - UMHS 0022 & 2023 \\
\href{http://doi.org/10.26275/ngob-yc3c}{10.26275/ngob-yc3c} & Stacey, W. & Michigan High Res EEG - UMHS 0019 & 2023 \\
\href{http://doi.org/10.26275/hhnz-div3}{10.26275/hhnz-div3} & Taylor-Clark, T. & Mapping of dorsal root ganglia sensory nerve populations in the mouse lung & 2023 \\
\href{http://doi.org/10.26275/jh8n-lwdl}{10.26275/jh8n-lwdl} & Danziger, Z. & Reflex voiding in rat occurs at consistent bladder volume regardless of pressure or infusion rate & 2023 \\
\href{http://doi.org/10.26275/ile7-wrsk}{10.26275/ile7-wrsk} & Bedard, P. & Deep phenotyping of Post-infectious Myalgic Encephalomyelitis-Chronic Fatigue Syndrome & 2023 \\
\href{http://doi.org/10.26275/eefp-azay}{10.26275/eefp-azay} & Biscola, N. & High-throughput segmentation of rat unmyelinated axons by deep learning & 2023 \\
\href{http://doi.org/10.26275/4p8l-n5kt}{10.26275/4p8l-n5kt} & Pattnaik, A. & iEEG networks preprocessing study & 2023 \\
\href{http://doi.org/10.26275/uz3t-mppx}{10.26275/uz3t-mppx} & Patel, B. & Fecobionics study in fecal incontinence (FI) human subjects & 2023 \\
\href{http://doi.org/10.26275/ahqc-ldar}{10.26275/ahqc-ldar} & Verheyden, J. & Identification of lung innervating sensory neurons and their target specificity in mouse (3) & 2023 \\
\href{http://doi.org/10.26275/sp7z-ylun}{10.26275/sp7z-ylun} & Moss, A. & Molecular phenotype distribution of single rat intracardiac neurons & 2023 \\
\href{http://doi.org/10.26275/qonw-sqix}{10.26275/qonw-sqix} & Musselman, E. & Computational model of vagus nerve stimulation in mice with bipolar cuff electrode & 2023 \\
\href{http://doi.org/10.26275/7mdx-asxc}{10.26275/7mdx-asxc} & Pelot, N. & Histology-based computational models of implanted human cervical vagus nerve stimulation with the LivaNova helical cuff electrode & 2023 \\
\href{http://doi.org/10.26275/df7j-e48n}{10.26275/df7j-e48n} & Musselman, E. & Pig-specific computational models of monopolar vagus nerve stimulation with a six-contact cuff electrode & 2023 \\
\href{http://doi.org/10.26275/wglu-jiud}{10.26275/wglu-jiud} & Pelot, N. & Validated computational models predict vagus nerve stimulation thresholds in preclinical animals and humans & 2023 \\
\href{http://doi.org/10.26275/uco6-ktjq}{10.26275/uco6-ktjq} & Tohara, K. & Mouse genetic models to manipulate enterochromaffin cell activity - Murine Organoid ELISA & 2023 \\
\href{http://doi.org/10.26275/j48r-vn5s}{10.26275/j48r-vn5s} & Patel, B. & Pilot Fecobionics study in healthy human subjects & 2023 \\
\href{http://doi.org/10.26275/ekv2-ohrk}{10.26275/ekv2-ohrk} & Patel, B. & Fecobionics study in healthy human subjects & 2023 \\
\href{http://doi.org/10.26275/dm5l-tspb}{10.26275/dm5l-tspb} & Armitage, O. & Influence of vagus nerve stimulation on cardiac activity in pigs & 2023 \\
\href{http://doi.org/10.26275/iv1j-x1g4}{10.26275/iv1j-x1g4} & Brookes, S. & Characterization of projections of long interneurons in human colon & 2023 \\
\href{http://doi.org/10.26275/rmcz-jfoq}{10.26275/rmcz-jfoq} & Nguyen, D. & Anterograde tracing of spinal afferent innervation in flat-mounts of the rat stomach & 2023 \\
\href{http://doi.org/10.26275/ikfe-z0u7}{10.26275/ikfe-z0u7} & Makhamreh, A. & Pseudouridine Quantification Data & 2023 \\
\href{http://doi.org/10.26275/btt6-puqw}{10.26275/btt6-puqw} & Pace, J. & In Vivo Labeling and Detection of Circulating Tumor Cells in Mice Using OTL38 & 2023 \\
\href{http://doi.org/10.26275/d488-z2q8}{10.26275/d488-z2q8} & Verheyden, J. & Identification of lung innervating sensory neurons and their target specificity in mouse (2) & 2023 \\
\href{http://doi.org/10.26275/5xqz-0dpe}{10.26275/5xqz-0dpe} & Iavarone, E. & SPARC Metadata Editor Tutorial – Creating new SDS Datasets & 2023 \\
\href{http://doi.org/10.26275/poex-tria}{10.26275/poex-tria} & Iavarone, E. & QuiltedTutorials 2 -  Re-Sampling data for computational simulations & 2023 \\
\href{http://doi.org/10.26275/nuwh-mtsd}{10.26275/nuwh-mtsd} & Iavarone, E. & SPARClet - Interactive flatmaps & 2023 \\
\href{http://doi.org/10.26275/f1pu-do9y}{10.26275/f1pu-do9y} & Iavarone, E. & SPARC Metadata Editor – Tutorials & 2023 \\
\href{http://doi.org/10.26275/5yvu-tr0d}{10.26275/5yvu-tr0d} & Iavarone, E. & Cardiovascular system model for comprehending the acute cardiac effects of vagus nerve stimulation & 2023 \\
\href{http://doi.org/10.26275/3vqo-xux0}{10.26275/3vqo-xux0} & Lin, M. & Scaffold map - Mapping of human gastric enteroendocrine cells & 2023 \\
\href{http://doi.org/10.26275/q8wb-a9qe}{10.26275/q8wb-a9qe} & Ivich, F. & Ratiometric fluorescence sensing and quantification of circulating blood sodium sensors in mice in vivo & 2023 \\
\href{http://doi.org/10.26275/8zuc-gst1}{10.26275/8zuc-gst1} & Aguinaga, D. & Mid-lumbar (L3) epidural stimulation effects on bladder and external urethral sphincter in non-injured and chronically transected urethane-anesthetized rats & 2023 \\
\href{http://doi.org/10.26275/gfwu-pi0p}{10.26275/gfwu-pi0p} & Haberbusch, M. & MicroCT imaging of the fascicular structure in the porcine right and left cervical vagus nerve & 2023 \\
\href{http://doi.org/10.26275/swkm-bzmg}{10.26275/swkm-bzmg} & Gould, T. & Calcium imaging tension recording and pellet transit in mouse colon in response to stimulation of the pelvic nerve (PNS) & 2023 \\
\href{http://doi.org/10.26275/xnv0-gtsj}{10.26275/xnv0-gtsj} & Gould, T. & Optogenetic activation of nitrergic and cholinergic neurons of murine colonic myenteric plexus & 2023 \\
\href{http://doi.org/10.26275/g9qv-rl2d}{10.26275/g9qv-rl2d} & Iavarone, E. & QuiltedTutorials 1 – Mapping 2D Data to 3D Organ Scaffold & 2023 \\
\href{http://doi.org/10.26275/efbj-8evl}{10.26275/efbj-8evl} & Blanz, S. & Stimulation of the pig vagus nerve to modulate target effect versus side effect & 2023 \\
\href{http://doi.org/10.26275/litx-swak}{10.26275/litx-swak} & Lin, M. & Scaffold map - Quantification of rat gastric enteroendocrine cells & 2023 \\
\href{http://doi.org/10.26275/puzi-xtm3}{10.26275/puzi-xtm3} & Brookes, S. & Antibodies tested in the colon – Human & 2023 \\
\href{http://doi.org/10.26275/aqri-vyb4}{10.26275/aqri-vyb4} & Brookes, S. & Targets of sympathetic nerves in myenteric plexus of human colon & 2023 \\
\href{http://doi.org/10.26275/rhda-nblc}{10.26275/rhda-nblc} & Yuan, P.-Q. & CLARITY and 3D imaging with high resolution and deep scanning of innervation in the pig colon by using SP8 DIVE fully tunable spectral multiphoton microscope & 2023 \\
\href{http://doi.org/10.26275/pa9i-likc}{10.26275/pa9i-likc} & Michel, K. & Ussing chamber experiments for distension evoked secretion in human colon & 2023 \\
\href{http://doi.org/10.26275/3q5o-leo5}{10.26275/3q5o-leo5} & Phillips, R. & Effect of chronic gastric electrical stimulation on the feeding behavior of diet-induced obese male Sprague-Dawley rats consuming a 45\% high-fat diet & 2023 \\
\href{http://doi.org/10.26275/ryft-516s}{10.26275/ryft-516s} & Dinning, P. & High resolution manometry & 2023 \\
\href{http://doi.org/10.26275/qjpu-t0sy}{10.26275/qjpu-t0sy} & Lin, M. & Scaffold map - Quantitative analysis of enteric neurons containing choline acetyltransferase and nitric oxide synthase immunoreactivities in the submucosal and myenteric plexuses of the porcine colon & 2023 \\
\href{http://doi.org/10.26275/n3uf-5doz}{10.26275/n3uf-5doz} & Larauche, M. & Influence of acute celiac branch of abdominal vagus nerve stimulation on colonic motility in anesthetized male Yucatan minipigs & 2023 \\
\href{http://doi.org/10.26275/abac-rzbv}{10.26275/abac-rzbv} & Larauche, M. & Influence of acute thoracolumbar root nerves electrical stimulation on colonic motility in anesthetized male Yucatan minipigs & 2023 \\
\href{http://doi.org/10.26275/6lqa-ghyr}{10.26275/6lqa-ghyr} & Tache, Y. & Influence of direct colon tissue electrical stimulation on colonic motility in anesthetized male Yucatan minipig & 2023 \\
\href{http://doi.org/10.26275/d41u-sokg}{10.26275/d41u-sokg} & Patel, B. & Triaxial mechanical testing of dog colon segments & 2023 \\
\href{http://doi.org/10.26275/wabs-qul8}{10.26275/wabs-qul8} & Vadigepalli, R. & Closed-loop modeling of central and intrinsic cardiac nervous system circuits underlying cardiovascular control & 2023 \\
\href{http://doi.org/10.26275/gc2e-qd0w}{10.26275/gc2e-qd0w} & Pattnaik, A. & Seizure severity score & 2023 \\
\href{http://doi.org/10.26275/zpxh-rqlw}{10.26275/zpxh-rqlw} & Iavarone, E. & BIOS Health Vagus Nerve Stimulation Calibrator & 2023 \\
\href{http://doi.org/10.26275/cdjq-pjlb}{10.26275/cdjq-pjlb} & Iavarone, E. & A multiscale predictive digital twin for neurocardiac modulation in rabbits & 2023 \\
\href{http://doi.org/10.26275/6b53-usyr}{10.26275/6b53-usyr} & Ardell, J. & RNA sequencing reveals novel transcripts from sympathetic stellate ganglia during cardiac sympathetic hyperactivity in rats & 2023 \\
\href{http://doi.org/10.26275/fi87-3n6o}{10.26275/fi87-3n6o} & Stebbing, M. & Distribution and coexpression patterns of specific cell markers of enteroendocrine cells in pig gastric epithelium & 2023 \\
\href{http://doi.org/10.26275/59t4-jlnz}{10.26275/59t4-jlnz} & Upadhye, A. & Human cervical vagus nerve fascicle imaging with MicroCT & 2023 \\
\href{http://doi.org/10.26275/dhe4-zh53}{10.26275/dhe4-zh53} & Iavarone, E. & Multi-scale human cardiac electrophysiology models & 2023 \\
\href{http://doi.org/10.26275/tfdf-w3hq}{10.26275/tfdf-w3hq} & Iavarone, E. & Expression Data Visualization tool & 2023 \\
\href{http://doi.org/10.26275/s00b-gket}{10.26275/s00b-gket} & Iavarone, E. & Multi-scale rabbit cardiac electrophysiology models & 2023 \\
\href{http://doi.org/10.26275/f5se-ynpk}{10.26275/f5se-ynpk} & Settell, M. & Histology of pig cervical vagus nerve & 2023 \\
\href{http://doi.org/10.26275/do5j-mz5q}{10.26275/do5j-mz5q} & Rodrigues, A. & Decoding vagus nerve activity with carbon nanotube sensors in freely moving rodents & 2023 \\
\href{http://doi.org/10.26275/zsop-bygv}{10.26275/zsop-bygv} & Bayrer, J. & Visceromotor responses (VMR) to colorectal distension in mice with silenced or activated enterochromaffin cells & 2023 \\
\href{http://doi.org/10.26275/wh9h-tbew}{10.26275/wh9h-tbew} & Bizanti, A. & Topographical mapping of sympathetic postganglionic innervation of the mouse heart & 2023 \\
\href{http://doi.org/10.26275/53cg-hai1}{10.26275/53cg-hai1} & Patel, B. & Computational-based mechanical analysis of Fecobionics data & 2023 \\
\href{http://doi.org/10.26275/yh5c-5pjy}{10.26275/yh5c-5pjy} & Cheng, Z. & Comparison of the intrinsic cardiac nervous system across male and female rat hearts & 2023 \\
\href{http://doi.org/10.26275/lk0e-kgrs}{10.26275/lk0e-kgrs} & Pelot, N. & Population of mock morphological models of vagus nerve stimulation with cuff electrodes for the purpose of studying the effect of fascicle diameter on activation threshold & 2023 \\
\href{http://doi.org/10.26275/ncok-1cof}{10.26275/ncok-1cof} & Pelot, N. & Computational model of heart rate modulation in mice during vagus nerve stimulation & 2023 \\
\href{http://doi.org/10.26275/kj2u-2ukf}{10.26275/kj2u-2ukf} & Gould, T. & Chemogenetic activation or inhibition of cholinergic or nitrergic myenteric neurons of mouse colon & 2023 \\
\href{http://doi.org/10.26275/nomg-p2vk}{10.26275/nomg-p2vk} & Ardell, J. & Cardioneural recordings using floating multi-channel plunge micro-electrodes in pigs & 2023 \\
\href{http://doi.org/10.26275/e3n7-vei4}{10.26275/e3n7-vei4} & Pelot, N. & Computational model of laryngeal muscle activation in mice during vagus nerve stimulation & 2023 \\
\href{http://doi.org/10.26275/sbxk-fwbu}{10.26275/sbxk-fwbu} & Iavarone, E. & Kember Cardiac Nerve Model & 2023 \\
\href{http://doi.org/10.26275/srm7-no3j}{10.26275/srm7-no3j} & Damaser, M. & Acute wired Urological Monitor of Conscious Activity (UroMOCA) implantation in feline bladder & 2023 \\
\href{http://doi.org/10.26275/6f3g-wvzh}{10.26275/6f3g-wvzh} & Damaser, M. & Acute Wired Colonic Monitor of Conscious Activity (ColoMOCA) implantation in pig bowel & 2023 \\
\href{http://doi.org/10.26275/z1wa-spub}{10.26275/z1wa-spub} & Hoover, D. & Regional analysis of autonomic nerves in normal and diseased human hearts & 2023 \\
\href{http://doi.org/10.26275/ggj4-agvt}{10.26275/ggj4-agvt} & Smith-Edwards, K. & Myenteric neuron activity during spontaneous motor complexes in mouse colon & 2023 \\
\href{http://doi.org/10.26275/kgkj-6vb9}{10.26275/kgkj-6vb9} & Feng, B. & Calcium imaging of mouse dorsal root ganglion (DRG) neurons in response to chemical stimuli of distal colon and rectum (colorectum) & 2023 \\
\href{http://doi.org/10.26275/dfk5-6w3z}{10.26275/dfk5-6w3z} & Horn, C. & Selective stimulation of the ferret abdominal vagus nerve with multi‐contact nerve cuff electrodes & 2023 \\
\href{http://doi.org/10.26275/d3jb-pkzx}{10.26275/d3jb-pkzx} & Harvey, R. & Sympathetic and parasympathetic effects on action potentials in isolated pig ventricular myocytes & 2023 \\
\href{http://doi.org/10.26275/lck3-f9v0}{10.26275/lck3-f9v0} & Taylor-Clark, T. & Mapping of the vagal afferent innervation of the mouse lung & 2023 \\
\href{http://doi.org/10.26275/tpu4-kvzo}{10.26275/tpu4-kvzo} & Jaffey, D. & Spatial distribution and morphometric characterization of mucosal afferents of the pylorus of the rat stomach & 2023 \\
\href{http://doi.org/10.26275/x1ht-tsiz}{10.26275/x1ht-tsiz} & Stebbing, M. & Innervation of enteroendocrine cells in the gastric mucosa in human and pig - including a description of the innervation of mucosal vasculature & 2023 \\
\href{http://doi.org/10.26275/f4p0-d1ic}{10.26275/f4p0-d1ic} & Rossen, N. & Mouse genetic models to manipulate enterochromaffin cell activity & 2023 \\
\href{http://doi.org/10.26275/h4yk-volr}{10.26275/h4yk-volr} & Gould, T. & Calcium imaging and tension recording in response to stimulation of the vagus nerve (VNS) & 2023 \\
\href{http://doi.org/10.26275/bcrr-uenq}{10.26275/bcrr-uenq} & Garg, R. & Defining neuromechanical mechanisms of Achilles tendinopathy progression- healthy human subjects (pilot) & 2022 \\
\href{http://doi.org/10.26275/vsez-uc77}{10.26275/vsez-uc77} & Ekhlasi, Z. & Generic rat brainstem scaffold & 2022 \\
\href{http://doi.org/10.26275/xcot-iplx}{10.26275/xcot-iplx} & Christie, R. & Generic pig heart scaffold & 2022 \\
\href{http://doi.org/10.26275/zfpg-gxgg}{10.26275/zfpg-gxgg} & Soltani, E. & Pig whole-body with embedded organs using automatic workflow for inserting the organs & 2022 \\
\href{http://doi.org/10.26275/g8aq-rjsp}{10.26275/g8aq-rjsp} & Brookes, S. & Characterization of projections of longitudinal muscle motor neurons in human colon & 2022 \\
\href{http://doi.org/10.26275/ueic-1afx}{10.26275/ueic-1afx} & Soltani, E. & Rat whole-body with embedded organs & 2022 \\
\href{http://doi.org/10.26275/zkol-rciv}{10.26275/zkol-rciv} & Soltani, E. & Rat whole-body with embedded organs using automatic workflow for inserting the organs & 2022 \\
\href{http://doi.org/10.26275/ewog-btml}{10.26275/ewog-btml} & Soltani, E. & Mouse whole-body with embedded organs & 2022 \\
\href{http://doi.org/10.26275/ffnm-sxbj}{10.26275/ffnm-sxbj} & Soltani, E. & Rat whole-body scaffold & 2022 \\
\href{http://doi.org/10.26275/5mkx-apz9}{10.26275/5mkx-apz9} & Soltani, E. & Human whole-body with embedded organs & 2022 \\
\href{http://doi.org/10.26275/biqn-mqy4}{10.26275/biqn-mqy4} & Smith-Edwards, K. & Enteric neuron responses in mouse distal colon to lumbosacral spinal cord stimulation & 2022 \\
\href{http://doi.org/10.26275/4iqi-hazf}{10.26275/4iqi-hazf} & Ludwig, K. & Anatomy and histology of the domestic pig in the context of vagus nerve stimulation & 2022 \\
\href{http://doi.org/10.26275/ltks-9kaw}{10.26275/ltks-9kaw} & Ekhlasi, Z. & Generic mouse brainstem scaffold & 2022 \\
\href{http://doi.org/10.26275/hmwa-nqdu}{10.26275/hmwa-nqdu} & Aristovich, K. & Organotopic organization of the porcine vagus nerve & 2022 \\
\href{http://doi.org/10.26275/ujsw-77cr}{10.26275/ujsw-77cr} & Christie, R. & Generic mouse heart scaffold & 2022 \\
\href{http://doi.org/10.26275/ajhr-1eom}{10.26275/ajhr-1eom} & Ekhlasi, Z. & Generic pig brainstem scaffold & 2022 \\
\href{http://doi.org/10.26275/0kpx-xh0x}{10.26275/0kpx-xh0x} & Christie, R. & Generic rat heart scaffold & 2022 \\
\href{http://doi.org/10.26275/hwgn-8mq4}{10.26275/hwgn-8mq4} & Ekhlasi, Z. & Generic sheep brainstem scaffold & 2022 \\
\href{http://doi.org/10.26275/pgs1-xfs5}{10.26275/pgs1-xfs5} & Christie, R. & Generic human heart scaffold & 2022 \\
\href{http://doi.org/10.26275/ezuv-o6gg}{10.26275/ezuv-o6gg} & Ekhlasi, Z. & Generic human brainstem scaffold & 2022 \\
\href{http://doi.org/10.26275/qmg0-zbde}{10.26275/qmg0-zbde} & Gould, T. & Calcium imaging tension recordings and pelet transit in mouse colon in response to direct electrical field stimulation  & 2022 \\
\href{http://doi.org/10.26275/eikd-dg5c}{10.26275/eikd-dg5c} & Yuan, P.-Q. & RNA sequencing analysis of transcriptomic responses to vagal nerve stimulation in myenteric ganglia of porcine colon & 2022 \\
\href{http://doi.org/10.26275/2yv6-g6vh}{10.26275/2yv6-g6vh} & Stacey, W. & Network dynamics of HFOs & 2022 \\
\href{http://doi.org/10.26275/hbuu-caud}{10.26275/hbuu-caud} & Gaunt, R. & Lower urinary tract nerve responses to high-density epidural spinal cord stimulation in cats & 2022 \\
\href{http://doi.org/10.26275/05g0-faru}{10.26275/05g0-faru} & Iavarone, E. & SPARC Metadata Editor (sparc-me) – Tutorials & 2022 \\
\href{http://doi.org/10.26275/yjtg-qxem}{10.26275/yjtg-qxem} & Iavarone, E. & SPARC Metadata Editor (sparc-me) - Creating new SDS datasets & 2022 \\
\href{http://doi.org/10.26275/pbrd-ci3l}{10.26275/pbrd-ci3l} & Ardell, J. & Identification of peripheral neural circuits that regulate heart rate using optogenetic and viral vector strategies & 2022 \\
\href{http://doi.org/10.26275/wcwr-hmch}{10.26275/wcwr-hmch} & Damaser, M. & Chronic wireless Urological Monitor of Conscious Activity (UroMOCA) implantation in feline bladder & 2022 \\
\href{http://doi.org/10.26275/iefx-c2qi}{10.26275/iefx-c2qi} & Lin, M. & Generic rat stomach scaffold & 2022 \\
\href{http://doi.org/10.26275/jnng-wbke}{10.26275/jnng-wbke} & Ekhlasi, Z. & Generic rat lung scaffold & 2022 \\
\href{http://doi.org/10.26275/zfbf-g88t}{10.26275/zfbf-g88t} & Lin, M. & Generic pig stomach scaffold & 2022 \\
\href{http://doi.org/10.26275/ugeq-kjky}{10.26275/ugeq-kjky} & Ekhlasi, Z. & Generic pig lung scaffold & 2022 \\
\href{http://doi.org/10.26275/in3y-t0ml}{10.26275/in3y-t0ml} & Lin, M. & Generic pig colon scaffold & 2022 \\
\href{http://doi.org/10.26275/u8bh-s6be}{10.26275/u8bh-s6be} & Lin, M. & Generic mouse stomach scaffold & 2022 \\
\href{http://doi.org/10.26275/nlfj-rhfk}{10.26275/nlfj-rhfk} & Ekhlasi, Z. & Generic mouse lung scaffold & 2022 \\
\href{http://doi.org/10.26275/z1ga-dflp}{10.26275/z1ga-dflp} & Lin, M. & Generic mouse colon scaffold & 2022 \\
\href{http://doi.org/10.26275/aijr-zl44}{10.26275/aijr-zl44} & Ekhlasi, Z. & Generic human lung scaffold & 2022 \\
\href{http://doi.org/10.26275/uhxm-wxpo}{10.26275/uhxm-wxpo} & Lin, M. & Generic human esophagus scaffold & 2022 \\
\href{http://doi.org/10.26275/7ph7-1ez4}{10.26275/7ph7-1ez4} & Lin, M. & Generic human colon scaffold & 2022 \\
\href{http://doi.org/10.26275/mxnk-4tvt}{10.26275/mxnk-4tvt} & Lin, M. & Scaffold map - Influence of direct colon tissue electrical stimulation on colonic motility in anesthetized male Yucatan minipig & 2022 \\
\href{http://doi.org/10.26275/amsj-jner}{10.26275/amsj-jner} & Pace, J. & Design and Validation of Near-Infared Diffuse In Vivo Flow Cytometry & 2022 \\
\href{http://doi.org/10.26275/umgm-rzar}{10.26275/umgm-rzar} & Howard, M. & Correlated electrophysiological immunohistochemical and morphological properties of proximal colon myenteric neurons & 2022 \\
\href{http://doi.org/10.26275/ebql-cdno}{10.26275/ebql-cdno} & Phillips, R. & Effect of chronic gastric electrical stimulation on the feeding behavior of male rats consuming a 45\% high-fat diet & 2022 \\
\href{http://doi.org/10.26275/spcl-epsl}{10.26275/spcl-epsl} & Liu, Z. & Effects of vagal afferent blockade on gastric motility during cervical vagus nerve stimulation measured with magnetic resonance imaging in rats & 2022 \\
\href{http://doi.org/10.26275/adzn-2fes}{10.26275/adzn-2fes} & Liu, Z. & In vivo mapping of gastric electrical activation with manganese enhanced magnetic resonance imaging & 2022 \\
\href{http://doi.org/10.26275/r5gw-clgv}{10.26275/r5gw-clgv} & Liu, Z. & Effects of vagal efferent blockade on gastric motility and emptying during cervical vagus nerve stimulation measured with magnetic resonance imaging in rats & 2022 \\
\href{http://doi.org/10.26275/y4k2-mkam}{10.26275/y4k2-mkam} & Phillips, R. & Effect of chronic gastric electrical stimulation on the feeding behavior of female rats consuming a 45\% high-fat diet & 2022 \\
\href{http://doi.org/10.26275/z4wa-dxjx}{10.26275/z4wa-dxjx} & Liu, Z. & Acute effects of gastric electrical stimulation settings on neural activity in the nucleus of solitary tract in rats & 2022 \\
\href{http://doi.org/10.26275/yt5s-pt6t}{10.26275/yt5s-pt6t} & Keast, J. & Central terminal fields of lower urinary tract afferents in rat & 2022 \\
\href{http://doi.org/10.26275/eedx-wrhi}{10.26275/eedx-wrhi} & Zeltser, L. & Expression of molecular markers in subpopulations of mouse superior cervical ganglion neurons  & 2022 \\
\href{http://doi.org/10.26275/s3iw-km0k}{10.26275/s3iw-km0k} & Zeltser, L. & Expression of molecular markers in subpopulations of mouse celiac ganglion neurons  & 2022 \\
\href{http://doi.org/10.26275/pfjd-cfty}{10.26275/pfjd-cfty} & Ardell, J. & Optical mapping of action potentials and calcium transients in the mouse heart during optogenetic stimulation of the intracardiac ganglia and interconnecting neurons (ICNS) & 2022 \\
\href{http://doi.org/10.26275/bvpu-cuz7}{10.26275/bvpu-cuz7} & Zeltser, L. & Single cell RNA sequencing of retrogradely labeled mouse stellate ganglion neurons & 2022 \\
\href{http://doi.org/10.26275/vm1h-k4kq}{10.26275/vm1h-k4kq} & Ludwig, K. & Electrode design characterization for electrophysiology from swine peripheral nervous system & 2022 \\
\href{http://doi.org/10.26275/zly9-ow0w}{10.26275/zly9-ow0w} & Zeltser, L. & Expression of molecular markers in subpopulations of mouse stellate ganglion neurons  & 2022 \\
\href{http://doi.org/10.26275/vkvk-3hqe}{10.26275/vkvk-3hqe} & Zeltser, L. & Expression of molecular markers in mouse and human stellate ganglia & 2022 \\
\href{http://doi.org/10.26275/23je-ute3}{10.26275/23je-ute3} & Jaffey, D. & 4D upper gastrointestinal magnetic resonance imaging in healthy human subjects and gastroparetic patients & 2022 \\
\href{http://doi.org/10.26275/dtlj-t4bf}{10.26275/dtlj-t4bf} & Tompkins, J. & Increased arrhythmia susceptibility in type 2 diabetic mice related to dysregulation of ventricular sympathetic innervation & 2022 \\
\href{http://doi.org/10.26275/jrj4-zrmj}{10.26275/jrj4-zrmj} & Nguyen, D. & Substance P-immunoreactive axon innervation of mouse stomach & 2022 \\
\href{http://doi.org/10.26275/akxk-vcbm}{10.26275/akxk-vcbm} & Rangel, A. & Electrochemical measurement of kanamycin in whole blood for implant longevity evaluation & 2022 \\
\href{http://doi.org/10.26275/jr6t-jcoe}{10.26275/jr6t-jcoe} & Gregersen, H. & Performance testing of the Fecobionics device & 2022 \\
\href{http://doi.org/10.26275/wjll-mekr}{10.26275/wjll-mekr} & Soltani, E. & Human whole-body with embedded organs using automatic workflow for inserting the organs & 2022 \\
\href{http://doi.org/10.26275/qq9z-e96l}{10.26275/qq9z-e96l} & Soltani, E. & Human whole-body scaffold & 2022 \\
\href{http://doi.org/10.26275/oujt-ao7w}{10.26275/oujt-ao7w} & Ekhlasi, Z. & Generic rat bladder scaffold & 2022 \\
\href{http://doi.org/10.26275/i6uj-vhx8}{10.26275/i6uj-vhx8} & Ludwig, K. & In vivo visualization of pig vagus nerve 'vagotopy' using ultrasound & 2022 \\
\href{http://doi.org/10.26275/tscp-vk1d}{10.26275/tscp-vk1d} & Gould, T. & Activity of and neurochemical expression in nitrergic and cholinergic neurons in the murine colonic myenteric plexus & 2022 \\
\href{http://doi.org/10.26275/ya41-6sfo}{10.26275/ya41-6sfo} & Feng, B. & Dorsal root ganglion stimulation to modulate mechanosensitive colorectal afferent transmission & 2022 \\
\href{http://doi.org/10.26275/iwxw-y2tz}{10.26275/iwxw-y2tz} & Liu, Z. & Assessment of gastric emptying and motility with magnetic resonance imaging (MRI) under gastric electrical stimulation (GES) in rats & 2022 \\
\href{http://doi.org/10.26275/36ua-upkq}{10.26275/36ua-upkq} & Feng, B. & Dorsal root ganglion stimulation to modulate mechanosensitive colorectal afferent transmission in mice & 2022 \\
\href{http://doi.org/10.26275/x0oc-7oc5}{10.26275/x0oc-7oc5} & Gould, T. & Calcium imaging and motility tracking of distinct myenteric neuronal subsets in mice & 2022 \\
\href{http://doi.org/10.26275/luv2-omyk}{10.26275/luv2-omyk} & Balachandran, K. & Generic pig bladder scaffold & 2022 \\
\href{http://doi.org/10.26275/14qt-e5f0}{10.26275/14qt-e5f0} & Ekhlasi, Z. & Generic cat bladder scaffold & 2022 \\
\href{http://doi.org/10.26275/mwdf-paqm}{10.26275/mwdf-paqm} & Balachandran, K. & Generic mouse bladder scaffold & 2022 \\
\href{http://doi.org/10.26275/qthf-0lou}{10.26275/qthf-0lou} & Ekhlasi, Z. & Generic human bladder scaffold & 2022 \\
\href{http://doi.org/10.26275/wcli-rv5b}{10.26275/wcli-rv5b} & Ardell, J. & Evaluating spheres of influence for efferent neural control of the heart & 2022 \\
\href{http://doi.org/10.26275/iyto-oxay}{10.26275/iyto-oxay} & Brierley, S. & Imaging colon and bladder sensory convergence in CLARITY cleared mouse spinal cord  & 2022 \\
\href{http://doi.org/10.26275/be0x-9mzy}{10.26275/be0x-9mzy} & Brierley, S. & Mapping colon and bladder innervating sensory neurons in CLARITY cleared ganglia in mouse & 2022 \\
\href{http://doi.org/10.26275/4mfy-y7bj}{10.26275/4mfy-y7bj} & Aristovich, K. & Temporal dispersion in porcine subdiaphragmatic nerves ex vivo & 2022 \\
\href{http://doi.org/10.26275/rmkt-5ypu}{10.26275/rmkt-5ypu} & Hanna, P. & Ablation of the intrinsic cardiac nervous system to evaluate efferent control of cardiac function & 2022 \\
\href{http://doi.org/10.26275/vzxw-kwdu}{10.26275/vzxw-kwdu} & Bruns, T. & High-density penetrating array dorsal root ganglia recordings & 2022 \\
\href{http://doi.org/10.26275/x10i-9c9u}{10.26275/x10i-9c9u} & Pelot, N. & Human vagus nerve TH- (tyrosine hydroxylase) and ChAT- (choline acetyltransferase) positive fibers & 2022 \\
\href{http://doi.org/10.26275/nav5-oeol}{10.26275/nav5-oeol} & Pelot, N. & Rat vagus nerve TH- (tyrosine hydroxylase) and ChAT- (choline acetyltransferase) positive fibers & 2022 \\
\href{http://doi.org/10.26275/zuwb-qnqk}{10.26275/zuwb-qnqk} & Howard, M. & Calcium dynamics imaging of vasoactive intestinal peptide-expressing (VIP) enteric nervous system (ENS) neurons & 2022 \\
\href{http://doi.org/10.26275/qw1u-zxea}{10.26275/qw1u-zxea} & Nicolai, E. & Vagus nerve stimulation mapping in swine & 2022 \\
\href{http://doi.org/10.26275/jsaw-2w1q}{10.26275/jsaw-2w1q} & Harvey, R. & Sympathetic and parasympathetic effects on membrane currents in isolated pig ventricular myocytes & 2022 \\
\href{http://doi.org/10.26275/1uno-tynt}{10.26275/1uno-tynt} & Wang, L. & Multicolor adeno-associated virus sparse labeling and 3D digital tracing of enteric plexus in mouse proximal colon & 2022 \\
\href{http://doi.org/10.26275/sip4-ioyz}{10.26275/sip4-ioyz} & Yuan, P.-Q. & CLARITY and three-dimensional (3D) imaging of the mouse and porcine colonic innervation & 2022 \\
\href{http://doi.org/10.26275/rt8g-gu2v}{10.26275/rt8g-gu2v} & Havton, L. & Ultrastructural analysis of human vagus nerve & 2022 \\
\href{http://doi.org/10.26275/qskp-awpu}{10.26275/qskp-awpu} & Gould, T. & Optogenetic inhibition of nitrergic and cholinergic neurons of murine colonic myenteric plexus & 2022 \\
\href{http://doi.org/10.26275/az1n-uv7s}{10.26275/az1n-uv7s} & Moss, A. & Spatially tracked single-cell-scale RNAseq of porcine right atrial ganglionic plexus (RAGP) neurons & 2022 \\
\href{http://doi.org/10.26275/cxej-bm2v}{10.26275/cxej-bm2v} & Heuckeroth, R. & Enteric nervous system expression profiling by high throughput scRNA-sequencing in human and mouse colon  & 2022 \\
\href{http://doi.org/10.26275/elbl-3vxh}{10.26275/elbl-3vxh} & Phillips, R. & Effect of electrical stimulation of vagal afferent terminals located in the stomach muscle wall on feeding behavior & 2022 \\
\href{http://doi.org/10.26275/wucy-ljuk}{10.26275/wucy-ljuk} & Verheyden, J. & Single nucleus RNAseq of nodose ganglia in mice & 2022 \\
\href{http://doi.org/10.26275/dhbx-w17y}{10.26275/dhbx-w17y} & Gregersen, H. & Safety testing of predicate device for Fecobionics & 2022 \\
\href{http://doi.org/10.26275/of13-iokw}{10.26275/of13-iokw} & Yuan, P.-Q. & Antibodies tested in the colon - Pig & 2022 \\
\href{http://doi.org/10.26275/z6jn-j5tx}{10.26275/z6jn-j5tx} & Moss, A. & Transcriptional diversity of single neurons in the porcine right atrial ganglionic plexus (RAGP) & 2022 \\
\href{http://doi.org/10.26275/9ly6-az0n}{10.26275/9ly6-az0n} & Zhuang, K. & SPARC Nerve Activity Predictor (SNAP) & 2022 \\
\href{http://doi.org/10.26275/s8jn-wboz}{10.26275/s8jn-wboz} & Zhuang, K. & Small Enteric Neural Network Simulator & 2022 \\
\href{http://doi.org/10.26275/oim6-urls}{10.26275/oim6-urls} & Zhuang, K. & 3D Visualization & 2022 \\
\href{http://doi.org/10.26275/jetk-pppi}{10.26275/jetk-pppi} & Zhuang, K. & 2D Plot & 2022 \\
\href{http://doi.org/10.26275/frbr-8uyt}{10.26275/frbr-8uyt} & Zhuang, K. & opencor-py model & 2022 \\
\href{http://doi.org/10.26275/gvzt-oeti}{10.26275/gvzt-oeti} & Khalifa, F. & Functional mapping with lumbosacral epidural stimulation for restoration of bladder function after spinal cord injury in rats & 2022 \\
\href{http://doi.org/10.26275/wlhm-c7ba}{10.26275/wlhm-c7ba} & Stanley, S. & Characterization of adeno associated virus serotypes 4 weeks after pancreas injection in mice & 2022 \\
\href{http://doi.org/10.26275/lok5-wje6}{10.26275/lok5-wje6} & Morotti, S. & Sympathetic nerve stimulation of mouse and rabbit hearts & 2022 \\
\href{http://doi.org/10.26275/ih2m-pphy}{10.26275/ih2m-pphy} & Liu, Z. & Acute effects of vagus nerve stimulation (VNS) settings on neural activity in the nucleus of solitary tract (NTS) in rats & 2022 \\
\href{http://doi.org/10.26275/by2v-qren}{10.26275/by2v-qren} & Zhuang, K. & Jupyterlab exploration of computational models of unmyelinated peripheral axons & 2022 \\
\href{http://doi.org/10.26275/9uqz-zwnh}{10.26275/9uqz-zwnh} & Liu, Z. & Acute effects of efferent and afferent vagus nerve stimulation (VNS) on neural activity accessed with functional Magnetic Resonance Imaging (fMRI) in rats & 2022 \\
\href{http://doi.org/10.26275/tbuz-s6gu}{10.26275/tbuz-s6gu} & Liu, Z. & Acute effects of gastric electrical stimulation settings on gastric motility assessed with magnetic resonance imaging in rats & 2022 \\
\href{http://doi.org/10.26275/g3xc-oztw}{10.26275/g3xc-oztw} & Sternini, C. & Quantitative analysis of enteric neurons containing choline acetyltransferase and nitric oxide synthase immunoreactivities in the submucosal and myenteric plexuses of the porcine colon & 2022 \\
\href{http://doi.org/10.26275/chfk-eugm}{10.26275/chfk-eugm} & Keast, J. & ViNERS (Visceral Nerve Ensemble Recording \& Stimulation) peripheral neural interface modeling environment & 2021 \\
\href{http://doi.org/10.26275/s2vo-pje2}{10.26275/s2vo-pje2} & Ardell, J. & Identification of peripheral neural circuits that regulate heart rate using optogenetic and viral vector strategies part (2) & 2021 \\
\href{http://doi.org/10.26275/kkmb-vun5}{10.26275/kkmb-vun5} & Keast, J. & Recording of electrically evoked neural activity and bladder pressure responses in awake rats chronically implanted with a pelvic nerve array & 2021 \\
\href{http://doi.org/10.26275/ts6z-z80x}{10.26275/ts6z-z80x} & Gregersen, H. & Safety testing of the Fecobionics device & 2021 \\
\href{http://doi.org/10.26275/e6vk-2lky}{10.26275/e6vk-2lky} & Verheyden, J. & Identification of lung innervating sensory neurons and their target specificity in mouse (1) & 2021 \\
\href{http://doi.org/10.26275/z61u-2tcs}{10.26275/z61u-2tcs} & Keast, J. & Simulations of pelvic and vagus neural interface anatomy-dependent stimulus and recording properties & 2021 \\
\href{http://doi.org/10.26275/ypwk-0xbo}{10.26275/ypwk-0xbo} & Liu, Z. & Acute effects of gastric electrical stimulation (GES) settings on neural activity accessed with functional magnetic resonance imaging (fMRI) in rats & 2021 \\
\href{http://doi.org/10.26275/ryap-anvh}{10.26275/ryap-anvh} & Ward, M. & An interactive system to visualize propagating vagal nerve activity in response to gastric electrical stimulation for gastroparesis & 2021 \\
\href{http://doi.org/10.26275/eoqv-ozxc}{10.26275/eoqv-ozxc} & Liu, Z. & Acute effects of efferent and afferent vagus nerve stimulation (VNS) on neural activity accessed with functional Magnetic Resonance Imaging (fMRI) in rats (Part 2) & 2021 \\
\href{http://doi.org/10.26275/fsfw-dmfd}{10.26275/fsfw-dmfd} & Chen, P. & Effects of subcutaneous nerve stimulation with blindly inserted electrodes on ventricular rate control in a canine model of persistent atrial fibrillation & 2021 \\
\href{http://doi.org/10.26275/c14g-2czn}{10.26275/c14g-2czn} & Moss, A. & Spatially tracked single-neuron transcriptomics of a male porcine right atrial ganglionic plexus (RAGP) & 2021 \\
\href{http://doi.org/10.26275/slsc-eahw}{10.26275/slsc-eahw} & Moss, A. & Spatially tracked single-neuron transcriptomics of a female porcine right atrial ganglionic plexus (RAGP) & 2021 \\
\href{http://doi.org/10.26275/hrww-enzr}{10.26275/hrww-enzr} & Moss, A. & Spatially tracked single-cell transcriptomics map of neuronal networks in the intrinsic cardiac nervous system & 2021 \\
\href{http://doi.org/10.26275/lzhy-ox20}{10.26275/lzhy-ox20} & Bernabei, J. & HUP Interictal iEEG Atlas & 2021 \\
\href{http://doi.org/10.26275/9ffg-482d}{10.26275/9ffg-482d} & Howard, M. & 3D imaging of enteric neurons in mouse & 2021 \\
\href{http://doi.org/10.26275/j5wc-rwcr}{10.26275/j5wc-rwcr} & Bruns, T. & Intraneural recordings in rat vagus nerves using carbon fiber microelectrode arrays & 2021 \\
\href{http://doi.org/10.26275/mhq6-csy1}{10.26275/mhq6-csy1} & Muenzberg, H. & iWAT (inguinal white adipose tissue) sympathetic innervation circuit pseudorabies viral tracing in reporter mice & 2021 \\
\href{http://doi.org/10.26275/drlw-j2cp}{10.26275/drlw-j2cp} & Zhuang, K. & Jupyter Voila Interactive Data Analysis & 2021 \\
\href{http://doi.org/10.26275/t8he-z5uu}{10.26275/t8he-z5uu} & Liu, Z. & Effects of nodose ganglion blockade on gastric motility during cervical vagus nerve stimulation measured with magnetic resonance imaging in rats & 2021 \\
\href{http://doi.org/10.26275/do5j-enxl}{10.26275/do5j-enxl} & Muenzberg, H. & Monosynaptic circuit mapping of iBAT (interscapular brown adipose tissues) in mice  & 2021 \\
\href{http://doi.org/10.26275/pb3l-251h}{10.26275/pb3l-251h} & Leung, C. & Mapping of intrinsic cardiac nervous system (ICN) neurons in a 3D reconstructed rat heart & 2021 \\
\href{http://doi.org/10.26275/pvib-4jat}{10.26275/pvib-4jat} & Muenzberg, H. & Endorgan-specific Pseudorabies (PRV) infection in mouse kidney and liver & 2021 \\
\href{http://doi.org/10.26275/ofja-ghoz}{10.26275/ofja-ghoz} & Pelot, N. & Quantified morphology of the human vagus nerve with anti-claudin-1 & 2021 \\
\href{http://doi.org/10.26275/er7m-gir3}{10.26275/er7m-gir3} & Khalifa, F. & Functional mapping with lumbosacral epidural stimulation for restoration of bladder function after spinal cord injury in rats (T13) & 2021 \\
\href{http://doi.org/10.26275/qcuk-a8ty}{10.26275/qcuk-a8ty} & Ludwig, K. & Sources of off-target effects for vagus nerve stimulation using the LivaNova clinical lead in swine & 2021 \\
\href{http://doi.org/10.26275/ngey-3iz7}{10.26275/ngey-3iz7} & Chen, P. & Effects of subcutaneous nerve stimulation on nerve sprouting in ambulatory dogs & 2021 \\
\href{http://doi.org/10.26275/wyn1-eww6}{10.26275/wyn1-eww6} & Howard, M. & 3D imaging of enteric neurons in a male mouse & 2021 \\
\href{http://doi.org/10.26275/lkvz-vrcy}{10.26275/lkvz-vrcy} & Yuan, P.-Q. & Single cell RNA sequencing (scRNAseq) analysis identifies the cell populations in the muscularis externa of the pig colon & 2021 \\
\href{http://doi.org/10.26275/ek1m-xqw1}{10.26275/ek1m-xqw1} & Harvey, R. & Sympathetic and parasympathetic effects on subcellular cAMP responses in isolated ventricular myocytes & 2021 \\
\href{http://doi.org/10.26275/i7dl-58h1}{10.26275/i7dl-58h1} & Wang, L. & Antibodies tested in the colon – Mouse & 2021 \\
\href{http://doi.org/10.26275/j4he-9spq}{10.26275/j4he-9spq} & Muenzberg, H. & Optogenetic iBAT (interscapular brown adipose tissue) stimulation in anesthetized mice & 2021 \\
\href{http://doi.org/10.26275/1h3s-thms}{10.26275/1h3s-thms} & Muenzberg, H. & Sympathetic iBAT (interscapular brown adipose tissue) activation high fat (HF) low fat (LF) diet study & 2021 \\
\href{http://doi.org/10.26275/jkux-orfg}{10.26275/jkux-orfg} & Fuller, D. & Phrenic nerve stimulation spinal intact rats & 2021 \\
\href{http://doi.org/10.26275/0khe-2os4}{10.26275/0khe-2os4} & Mazzuoli-Weber, G. & In vitro imaging of mechanosensitive submucous neurons in the porcine colon & 2021 \\
\href{http://doi.org/10.26275/imbg-0okx}{10.26275/imbg-0okx} & Damaser, M. & Effects of cystotomy on the feline urinary bladder & 2021 \\
\href{http://doi.org/10.26275/advv-1awo}{10.26275/advv-1awo} & Sladek, J. & Optogenetic stimulation prevents lipopolysaccharide induced TNFa production & 2021 \\
\href{http://doi.org/10.26275/ckgb-5ewo}{10.26275/ckgb-5ewo} & Muenzberg, H. & TRAP-SEQ (Translating Ribosome Affinity Purification followed by RNA sequencing) of interscapular brown adipose tissue (iBAT)- related ganglia from 7-day cold and warm treated mice & 2021 \\
\href{http://doi.org/10.26275/8vi1-tcsp}{10.26275/8vi1-tcsp} & Morris, K. & Tool for uploading files to Blackfynn & 2021 \\
\href{http://doi.org/10.26275/56h4-ypua}{10.26275/56h4-ypua} & Moss, A. & Acquisition of single neurons and regional neuronal samples from the porcine right atrial ganglionic plexus (RAGP) through laser capture microdissection & 2021 \\
\href{http://doi.org/10.26275/pkgd-bopz}{10.26275/pkgd-bopz} & Muenzberg, H. & Retrograde tracing of interscapular brown adipose tissue (iBAT) specific sympathetic neurons in mice - virus and reporter testing & 2021 \\
\href{http://doi.org/10.26275/m9ti-0pbj}{10.26275/m9ti-0pbj} & Muenzberg, H. & Gene expression profile of interscapular brown adipose tissue (iBAT) and inguinal white adipose tissue (iWAT) whole ganglia sequencing in mice & 2021 \\
\href{http://doi.org/10.26275/pidf-15l3}{10.26275/pidf-15l3} & Muenzberg, H. & Chemogenetic whole-body and iBAT (interscapular brown adipose tissue) -specific sympathetic stimulation in anesthetized mice & 2021 \\
\href{http://doi.org/10.26275/tuof-9odl}{10.26275/tuof-9odl} & Muenzberg, H. & Chemogenetic iBAT (interscapular brown adipose tissue)-specific sympathetic stimulation and e-mitter implant in mice & 2021 \\
\href{http://doi.org/10.26275/rtzw-x9u4}{10.26275/rtzw-x9u4} & Powley, T. & Functional mapping of the stomach neural circuitry - gastric electrical stimulation (GES) evoked duodenal motility in rats & 2021 \\
\href{http://doi.org/10.26275/zxe9-o3ss}{10.26275/zxe9-o3ss} & Powley, T. & MicroCT imaging of rat stomach vasculature with Microfil MV-122 & 2021 \\
\href{http://doi.org/10.26275/6xtv-zfpc}{10.26275/6xtv-zfpc} & Scheller, E. & Spatial mapping and contextualization of axon subtypes innervating the long bones of C3H and B6 mice & 2021 \\
\href{http://doi.org/10.26275/w027-cisv}{10.26275/w027-cisv} & Yang, H. & Imaging in vivo acetylcholine release in the peripheral nervous system with a fluorescent nanosensor in mice & 2021 \\
\href{http://doi.org/10.26275/jl5t-xfgu}{10.26275/jl5t-xfgu} & Powley, T. & Micro Computed Tomography (Micro-CT) imaging of iodine-stained rat stomachs from full to empty & 2021 \\
\href{http://doi.org/10.26275/gdot-t59p}{10.26275/gdot-t59p} & Keast, J. & Immunohistochemical classification of sensory and autonomic neurons projecting to the lower urinary tract in rats & 2021 \\
\href{http://doi.org/10.26275/fcrd-lbid}{10.26275/fcrd-lbid} & Campbell-Thompson, M. & Human islet microvasculature analysis & 2021 \\
\href{http://doi.org/10.26275/dap3-ckep}{10.26275/dap3-ckep} & Pelot, N. & Pig vagus nerve TH (tyrosine hydroxylase) and ChAT (choline acetyltransferase) positive fibers & 2021 \\
\href{http://doi.org/10.26275/bjp1-ppqo}{10.26275/bjp1-ppqo} & Vaseghi, M. & Functional neuronal nodose recording from pig - Cardiac field chemical and mechanical stimulation & 2020 \\
\href{http://doi.org/10.26275/mguq-j2n3}{10.26275/mguq-j2n3} & Nanivadekar, A. & Selectivity of afferent microstimulation at the DRG using epineural and penetrating electrode arrays & 2020 \\
\href{http://doi.org/10.26275/x9xq-e4wu}{10.26275/x9xq-e4wu} & Williams, A. & LLC and MM DiFC CTC detections 2020\_11\_04 & 2020 \\
\href{http://doi.org/10.26275/tv7g-o8ff}{10.26275/tv7g-o8ff} & Clancy, C. & A multi-scale model of cardiac electrophysiology & 2020 \\
\href{http://doi.org/10.26275/ilkm-9f8r}{10.26275/ilkm-9f8r} & Biscola, N. & Morphometric analysis of the abdominal vagus nerve in rats & 2020 \\
\href{http://doi.org/10.26275/maq2-eii4}{10.26275/maq2-eii4} & Pelot, N. & Quantified morphology of the pig vagus nerve  & 2020 \\
\href{http://doi.org/10.26275/sydt-lkiw}{10.26275/sydt-lkiw} & Pelot, N. & Human vagus nerve stained with Masson's trichrome & 2020 \\
\href{http://doi.org/10.26275/ilb9-0e2a}{10.26275/ilb9-0e2a} & Pelot, N. & Quantified morphology of the rat vagus nerve & 2020 \\
\href{http://doi.org/10.26275/8pc2-rhu2}{10.26275/8pc2-rhu2} & Pelot, N. & Quantified morphology of the pig vagus nerve with anti-fibronectin & 2020 \\
\href{http://doi.org/10.26275/jg3k-z5qm}{10.26275/jg3k-z5qm} & Keast, J. & Immediate early gene (IEG) mapping of spinal cord neurons activated by cystometry induced micturition in rat & 2020 \\
\href{http://doi.org/10.26275/0y4e-eskx}{10.26275/0y4e-eskx} & Gould, T. & Distribution of nitregic cholinergic and all  myenteric plexus neurons & 2020 \\
\href{http://doi.org/10.26275/k64z-n56w}{10.26275/k64z-n56w} & Wagenaar, J. & Scientific Articles & 2020 \\
\href{http://doi.org/10.26275/kw6v-vftb}{10.26275/kw6v-vftb} & Pascual, I. & Neurofauna Rat & 2020 \\
\href{http://doi.org/10.26275/boe7-1bms}{10.26275/boe7-1bms} & Horn, C. & Characterizing the effect of feeding distension and emetic stimuli on gastric myoelectric activity in ferrets & 2020 \\
\href{http://doi.org/10.26275/z3ab-7j9y}{10.26275/z3ab-7j9y} & Pelot, N. & Rat vagus nerve stained with Masson's trichrome & 2020 \\
\href{http://doi.org/10.26275/pgr9-bk2e}{10.26275/pgr9-bk2e} & Pelot, N. & Pig vagus nerve stained with Masson's trichrome & 2020 \\
\href{http://doi.org/10.26275/hgwy-pchm}{10.26275/hgwy-pchm} & Calderon, I. & Real-time particle-by-particle detection of erythrocyte camouflaged microsensor with extended circulation time in the bloodstream & 2020 \\
\href{http://doi.org/10.26275/owri-mpsx}{10.26275/owri-mpsx} & Ardell, J. & Functional recordings from the pig intrinsic cardiac nervous system (ICN) & 2020 \\
\href{http://doi.org/10.26275/nnyt-bqpg}{10.26275/nnyt-bqpg} & Mitchell, G. & Phrenic nerve immunohistochemistry & 2020 \\
\href{http://doi.org/10.26275/prjd-jhoc}{10.26275/prjd-jhoc} & Clark, H. & Submandibular ganglion stained by bungarotoxin and nanosensors in mouse & 2020 \\
\href{http://doi.org/10.26275/63lh-hdz5}{10.26275/63lh-hdz5} & Chen, P. & Electrophysiology in dog after subcutaneous nerve stimulation & 2020 \\
\href{http://doi.org/10.26275/xkoa-oqec}{10.26275/xkoa-oqec} & Muenzberg, H. & iBAT (interscapular Brown Adipose Tissue) sympathetic innervation circuit pseudorabies viral tracing in reporter mice & 2020 \\
\href{http://doi.org/10.26275/4qvr-kwzq}{10.26275/4qvr-kwzq} & Schemann, M. & Ussing chamber pressure pump & 2020 \\
\href{http://doi.org/10.26275/osy6-dn3o}{10.26275/osy6-dn3o} & Helmer, K. & Bilateral recordings of cervical vagus nerve activity in rats & 2020 \\
\href{http://doi.org/10.26275/t6j6-77pu}{10.26275/t6j6-77pu} & Wagenaar, J. & Hetionet & 2020 \\
\href{http://doi.org/10.26275/uztw-z5sc}{10.26275/uztw-z5sc} & Clancy, C. & Prototype simulation of undiseased human cardiac ventricular cells & 2020 \\
\href{http://doi.org/10.26275/1upo-xvkt}{10.26275/1upo-xvkt} & Morris, K. & Feline brainstem neuron extracellular potential recordings & 2020 \\
\href{http://doi.org/10.26275/ukz3-0fao}{10.26275/ukz3-0fao} & Powley, T. & Spatial distribution and morphometric characterization of vagal efferents associated with the myenteric plexus of the rat stomach & 2020 \\
\href{http://doi.org/10.26275/3m8n-0owa}{10.26275/3m8n-0owa} & Powley, T. & Spatial distribution and morphometric characterization of vagal afferents (intramuscular arrays (IMAs)) within the longitudinal and circular muscle layers of the rat stomach & 2020 \\
\href{http://doi.org/10.26275/wzry-sf7v}{10.26275/wzry-sf7v} & Powley, T. & Spatial distribution and morphometric characterization of vagal afferents associated with the myenteric plexus of the rat stomach & 2020 \\
\href{http://doi.org/10.26275/spfh-lx9g}{10.26275/spfh-lx9g} & Aristovich, K. & Imaging fast neural traffic at fascicular level with electrical impedance tomography - Proof of principle in rat sciatic nerve & 2020 \\
\href{http://doi.org/10.26275/o9qr-l4x9}{10.26275/o9qr-l4x9} & Stebbing, M. & Quantification of rat gastric enteroendocrine cells & 2020 \\
\href{http://doi.org/10.26275/mzth-oxbk}{10.26275/mzth-oxbk} & Stebbing, M. & Quantification of the relationship between rat gastric nerve fibers and enteroendocrine cells (EEC) & 2020 \\
\href{http://doi.org/10.26275/ppgj-qqpf}{10.26275/ppgj-qqpf} & Stebbing, M. & Mapping of human gastric enteroendocrine cells & 2020 \\
\href{http://doi.org/10.26275/w4my-puqm}{10.26275/w4my-puqm} & Ardell, J. & Functional neuronal nodose recording from pig- Modulation by myocardial ischemia and variably coupled PVC's & 2020 \\
\href{http://doi.org/10.26275/xmsp-wwtu}{10.26275/xmsp-wwtu} & Stanley, S. & Quantification of Cholera Toxin Subunit Beta (CTb) positive neurons in the coeliac nodose and dorsal root ganglia 1 week after pancreas injection in mice & 2020 \\
\href{http://doi.org/10.26275/dwzu-xtmj}{10.26275/dwzu-xtmj} & Zeltser, L. & Cholera toxin B retrograde tracing from brown adipose tissue and forelimb to the stellate ganglion & 2020 \\
\href{http://doi.org/10.26275/ge74-ypxd}{10.26275/ge74-ypxd} & Zeltser, L. & Visualizing sympathetic projections in the intact brown adipose tissue depot in the mouse & 2020 \\
\href{http://doi.org/10.26275/pzek-91wx}{10.26275/pzek-91wx} & Heuckeroth, R. & Robust 3-Dimensional visualization of human colon enteric nervous system without tissue sectioning & 2020 \\
\href{http://doi.org/10.26275/c4xq-9kl0}{10.26275/c4xq-9kl0} & Gonzalez-Rothi, E. & Effect of intermittent hypoxia preconditioning in rats with chronic cervical spinal cord injury – An electrophysiological study & 2020 \\
\href{http://doi.org/10.26275/mvwc-fnqm}{10.26275/mvwc-fnqm} & Powley, T. & Vagus nerve stimulation promotes gastric emptying by increasing pyloric opening measured with magnetic resonance imaging & 2020 \\
\href{http://doi.org/10.26275/guqw-r3ca}{10.26275/guqw-r3ca} & Helmer, K. & Processed fMRI data of transcutaneous auricular vagus nerve (taVNS) stimulation in humans & 2020 \\
\href{http://doi.org/10.26275/qh3q-elj6}{10.26275/qh3q-elj6} & Ward, M. & Influence of left vagal stimulus pulse parameters on vagal and gastric activity in rat & 2020 \\
\href{http://doi.org/10.26275/t4ng-2zm6}{10.26275/t4ng-2zm6} & McCallum, G. & Chronic interfacing with the autonomic nervous system using carbon nanotube (CNT) yarn electrodes & 2020 \\
\href{http://doi.org/10.26275/duz8-mq3n}{10.26275/duz8-mq3n} & Patel, B. & Computational modelling of the mechanical behavior of the colon & 2020 \\
\href{http://doi.org/10.26275/mq7u-tdjd}{10.26275/mq7u-tdjd} & Pascual, I. & Anatomical Model Viewer & 2019 \\
\href{http://doi.org/10.26275/pfxu-irqf}{10.26275/pfxu-irqf} & Johnson, J. & Intracranial EEG Epilepsy - Study 6 & 2019 \\
\href{http://doi.org/10.26275/psj7-wggf}{10.26275/psj7-wggf} & Johnson, J. & Intracranial EEG Epilepsy - Study 3 & 2019 \\
\href{http://doi.org/10.26275/hqyv-oagj}{10.26275/hqyv-oagj} & Johnson, J. & Intracranial EEG Epilepsy - Study 5 & 2019 \\
\href{http://doi.org/10.26275/ju7v-hnyy}{10.26275/ju7v-hnyy} & Boccanfuso, J. & Canine Epilepsy Dataset & 2019 \\
\bottomrule

\end{longtable}
\end{landscape}
\FloatBarrier

\clearpage
\section{Review of Neuroscience Platforms}

\begin{table}[ht]
\caption{Brain-CODE review.}
\label{tab:Brain-CODE}
\centering
\makebox[\textwidth][c]{
    \renewcommand{\arraystretch}{1.5}
    \begin{tabular}{>{\raggedright\arraybackslash}p{5.6cm} p{10cm}}
    \textbf{Data Management Opportunity} & \textbf{Comments}\\
    \hline 
    Multimodal Data Management & Supports numerous data types including clinical, neuroimaging, molecular, and multiomics data from humans, animal models, and basic science research. \\
    Comprehensive Metadata Support & Includes metadata on study protocols, equipment, and participant-related information. These inform data access controls and allow data to be combined across research programs. \\
    FAIR Data Sharing & The development of Brain-CODE is directly guided by the FAIR principles. It features data sharing policies, standardization efforts, and a federated data architecture. \\
    Optimizing Data Reliability and Utilization & Multiple data quality assurance and control processes are implemented. It provides tools for data capture and exploration, as well as workspaces with dedicated computing resources for processing and analytics. \\
    Facilitating Data Integration and Standardization & Implements CDEs, ontologies, and standardized processes across studies. It also applies the Clinical Data Interchange Standards Consortium (CDISC) standards where possible. \\
    Fostering Collaborative Science & Data federation and linkages with external databases support data sharing with collaborators. Workspace environments centralize data workflows. \\
    Enabling Scalable Analysis & Provides scalable computing resources and integrated tools for data analysis. \\
    Ensuring Resource Sustainability & Federation with national and international databases, a detailed governance structure, and an extensibly-designed platform help it's long-term utility. The platform is closed-source and hosted at a high-performance computing facility (Centre for Advanced Computing, Ontario, Canada).\\ 
    \hline
    \end{tabular}
}
\end{table}

\begin{table}[ht]
\caption{brainlife.io review.}
\label{tab:brainlife}
\centering
\makebox[\textwidth][c]{
    \renewcommand{\arraystretch}{1.5}
    \begin{tabular}{>{\raggedright\arraybackslash}p{5.6cm} p{10cm}}
    \textbf{Data Management Opportunity} & \textbf{Comments} \\
    \hline 
    Multimodal Data Management & Handles multiple neuroimaging modalities including MRI, MEG, and EEG data. While limited in modalities, it improves the interoperability of these data formats from different sources. \\
    Comprehensive Metadata Support & Although brainlife.io tracks data provenance and uses standardized data formats, there is not an explicit focus on comprehensive metadata management. \\
    FAIR Data Sharing & Developed specifically in support of the FAIR principles. \\
    Optimizing Data Reliability and Utilization & Focuses heavily on enabling reliable data processing, analysis, and reuse across multiple neuroimaging modalities through its analysis workflows and tools. \\
    Facilitating Data Integration and Standardization & Implements 'Datatypes' as a key standardization feature that make data interoperable across applications. \\
    Fostering Collaborative Science & Supports both public and private projects. Enables the sharing of apps and computational resources for collaborative research. \\
    Enabling Scalable Analysis & Integrates high-performance computing and cloud resources to support scalable analysis. \\
    Ensuring Resource Sustainability & Operates on a public-funding model and supports a large user base across numerous institutions. The platform is fully open-source and the cloud architecture is described in-detail. \\ 
    \hline
    \end{tabular}
}
\end{table}

\begin{table}[ht]
\caption{DABI review.}
\label{tab:DABI}
\centering
\makebox[\textwidth][c]{
    \renewcommand{\arraystretch}{1.5}
    \begin{tabular}{>{\raggedright\arraybackslash}p{5.6cm} p{10cm}}
    \textbf{Data Management Opportunity} & \textbf{Comments} \\
    \hline 
    Multimodal Data Management & Supports various data types including electrophysiology, clinical, imaging, pathology, demographics, and behavioral data. \\
    Comprehensive Metadata Support & Implements metadata management through data harmonization and CDEs. Metadata can be querying and visualized with the ability to create cohorts. \\
    FAIR Data Sharing & While not adhering to the FAIR principles explicitly, DABI implements practices that align them: accessible data sharing, standardized formats for data interoperability, and features that support data reuse. \\
    Optimizing Data Reliability and Utilization & Data provenance is captured and retained alongside the data. Additionally, an audit trail tracks changes made to datasets. Provides tools for data analysis, visualization, and quality control. \\
    Facilitating Data Integration and Standardization & Encourages the use of standardized formats like BIDS and NWB. Metadata files are required, and a metadata harmonization tool is provided for data in non-compliant formats. \\
    Fostering Collaborative Science & Collaborative features on the platform are limited to data sharing and access control. \\
    Enabling Scalable Analysis & Integrates with analytical tools, offers a machine learning ecosystem, and allows for data exploration without needing to download raw data. \\
    Ensuring Resource Sustainability & LONI aims to maintain the archive after funding ends without charging users fees for data access and has historically done so for other projects. The platform is built with a cloud architecture, although it is closed-source. \\ 
    \hline
    \end{tabular}
}
\end{table}

\begin{table}[ht]
\caption{DANDI review.}
\label{tab:DANDI}
\centering
\makebox[\textwidth][c]{
    \renewcommand{\arraystretch}{1.5}
    \begin{tabular}{>{\raggedright\arraybackslash}p{5.6cm} p{10cm}}
    \textbf{Data Management Opportunity} & \textbf{Comments} \\
    \hline 
    Multimodal Data Management & Accepts electrophysiology, optophysiology, behavioral time-series, and images from immunostaining experiments, among others. \\
    Comprehensive Metadata Support & Supports metadata, stores them together with data, and has a standard schema. Capabilities beyond organization are limited. \\
    FAIR Data Sharing & DANDI is compliant with the FAIR principles and prioritizes open access data. Assigns DOIs to dataset versions for citability. \\
    Optimizing Data Reliability and Utilization & Uploaded data are validated for adherence to standardized formats and it has data identifiers for versioning datasets and assets. Allows programmatic access to data in the cloud and provides built-in tools for organization, visualization, and analysis. \\
    Facilitating Data Integration and Standardization & Encourages the use of data standards (NWB, BIDS, NIDM, and other BRAIN Initiative standards) for consistency and interoperability across datasets. \\
    Fostering Collaborative Science & Projects on DANDI allow collaborators and the integration with JupyterHub potentially enables shared data analyses. Built-in, collaborative features on the platform itself are minimal. \\
    Enabling Scalable Analysis & Allows data streaming for access to parts of datasets. Has compute-near-data functionality through DANDI Hub, enabling users to perform analyses on large datasets without local downloads.\\
    Ensuring Resource Sustainability & A partnership with the BRAIN Initiative and funding from NIH support long-term sustainability. The platform itself is cloud-based and open-source.\\ 
    \hline
    \end{tabular}
}
\end{table}

\begin{table}[ht]
\caption{EBRAINS review.}
\label{tab:EBRAINS}
\centering
\makebox[\textwidth][c]{
    \renewcommand{\arraystretch}{1.5}
    \begin{tabular}{>{\raggedright\arraybackslash}p{5.6cm} p{10cm}}
    \textbf{Data Management Opportunity} & \textbf{Comments} \\
    \hline 
    Multimodal Data Management & Hosts data from all modalities and species, as well as models, software, and atlases. \\
    Comprehensive Metadata Support & A dedicated curation team supports researchers in describing and annotating their data. The platform uses the openMINDS metadata framework for standardized metadata management. \\
    FAIR Data Sharing & EBRAINS is committed to the FAIR principles — providing a platform with open access policies and capabilities for data integration and sharing. \\
    Optimizing Data Reliability and Utilization & Provides long-term storage solutions and emphasizes data reusability. Includes tools for modeling and simulation. \\
    Facilitating Data Integration and Standardization & Data is standardized through the curation process using community-driven metadata standards and ontologies. Consistent and precise organization is prioritized over strict standardization. \\
    Fostering Collaborative Science & Offers collaborative, cloud workspaces for private and community collaboration. Provides tools for sharing and accessing data, software, models, and atlases. \\
    Enabling Scalable Analysis & Provides access to high-performance computing resources for large-scale data analysis and simulations. \\
    Ensuring Resource Sustainability & Provides free data storage at the Swiss National Supercomputing Centre for at least 10 years. Substantial funding from the EU supports long-term sustainability and it is part of the European Open Science Cloud association (EOSC). EBRAINS has open-source elements, but the platform is not fully open-source. \\ 
    \hline
    \end{tabular}
}
\end{table}

\begin{table}[ht]
\caption{The IDA review.}
\label{tab:IDA}
\centering
\makebox[\textwidth][c]{
    \renewcommand{\arraystretch}{1.5}
    \begin{tabular}{>{\raggedright\arraybackslash}p{5.6cm} p{10cm}}
    \textbf{Data Management Opportunity} & \textbf{Comments} \\
    \hline 
    Multimodal Data Management & Manages various types of neuroscience data, including MRI, PET, SPECT, and EEG. Also supports multiomics, clinical, and biospecimen data, among other modalities. \\
    Comprehensive Metadata Support & While demographic and clinical metadata are included and searchable, there's not a comprehensive approach to metadata management. \\
    FAIR Data Sharing & The IDA implements practices that align with the FAIR principles including data archiving, exploring, and sharing. However, some FAIR strategies like dataset DOI assignment are not implemented. \\
    Optimizing Data Reliability and Utilization & It ensures reliability and uptime through fault-tolerant network infrastructure. Features advanced search capabilities to build data collections and easy download options.  \\
    Facilitating Data Integration and Standardization & Offers tools for de-identifying data and ensures compliance with patient-privacy regulations. It also provides standardized processes for data upload and download. However, there's no broader data standardization efforts across studies. \\
    Fostering Collaborative Science & Designed to facilitate data transfers between collaborative groups. Has an API and tools for external developers to access and integrate data on the IDA. However, collaborative features within the platform are lacking.\\
    Enabling Scalable Analysis & The IDA doesn't directly provide analysis tools, but the API allows external tools and workflows to access data. \\
    Ensuring Resource Sustainability & Widely used for the long-term preservation of neuroscience and biomedical research data. Although the platform is closed-source, its infrastructure is robust, and it is supported by major institutions. \\ 
    \hline
    \end{tabular}
}
\end{table}

\begin{table}[ht]
\caption{OpenNeuro review.}
\label{tab:openneuro}
\centering
\makebox[\textwidth][c]{
    \renewcommand{\arraystretch}{1.5}
    \begin{tabular}{>{\raggedright\arraybackslash}p{5.6cm} p{10cm}}
    \textbf{Data Management Opportunity} & \textbf{Comments} \\
    \hline 
    Multimodal Data Management & Supports MRI, EEG, iEEG, MEG, and PET data, among other BIDS-compatible data types. \\
    Comprehensive Metadata Support & Includes basic metadata through the BIDS format and additional dataset-level metadata can be provided by dataset contributors. While the extent of metadata features is limited by the BIDS specification, this ensures consistency across datasets. \\
    FAIR Data Sharing & Explicitly addresses each FAIR principle through persistent, unique identifiers, multiple data retrieval methods, BIDS standardization, and open data sharing. \\
    Optimizing Data Reliability and Utilization & Offers easy downloading options via web and programming interfaces. Data is well-organized and structured. Uses content-addressing, checksums, and incremental backups to ensure data integrity. \\
    Facilitating Data Integration and Standardization & Strictly adheres to BIDS format, ensuring a high level of standardization across datasets. \\
    Fostering Collaborative Science & Primarily facilitates collaboration through public data sharing. Direct collaborative features within data archives are not prioritized. \\
    Enabling Scalable Analysis & While OpenNeuro itself focuses on data sharing rather than analysis, cloud-based analysis and visualization tools are provided on partner platforms. \\
    Ensuring Resource Sustainability & Receives support and funding through the BRAIN Initiative and other large organizations. The platform is cloud-based and fully open-source. \\ 
    \hline
    \end{tabular}
}
\end{table}

\end{document}